\documentclass[a4paper,11pt]{article}
\pdfoutput=1
\usepackage{amsmath,amssymb,graphicx,todonotes,xspace,cite,hyperref,slashed,subcaption,float,tabularx}
\usepackage[section]{placeins}
\newcommand{\Ca}{\ensuremath{C_{\!A}}\xspace}

\newcommand{\as}{\ensuremath{\alpha_s}\xspace}
\newcommand{\HEJ}{{\tt HEJ}\xspace}

\newcommand{\HEJtwo}{{\tt HEJ2}\xspace}

\title{\begin{normalsize}
\begin{flushright}
DCPT/21/14, DESY 21-107, IPPP/21/07,\\MCnet-21-13, SAGEX-21-16
\end{flushright}
\end{normalsize}
\vspace*{2cm}Logarithmic corrections to the QCD component of\\ same-sign  W-pair
production for VBS studies}

\author{Jeppe R.~Andersen$^{a}$, Bertrand Duclou\'e$^{b}$, Conor Elrick$^{b}$,\\
  Andreas Maier$^{c}$, %
Graeme Nail$^{b}$, Jennifer M.~Smillie$^{b}$}
\date{%
$^{a}$ Institute for Particle Physics Phenomenology,\\ University of Durham,
Durham, DH1 3LE, UK\\%
$^{b}$ Higgs Centre for Theoretical Physics, University of Edinburgh,\\
  Peter Guthrie Tait Road, Edinburgh, EH9 3FD, UK\\%
$^{c}$ Deutsches Elektronen-Synchrotron DESY,\\ Platanenallee 6,
  15738 Zeuthen, Germany}

\begin{document}

\maketitle

\begin{abstract}
We present the results of the first calculation of the logarithmic
  corrections to the QCD contribution to same-sign $W$-pair
  production, $pp\to e^\pm \nu_e \mu^\pm
  \nu_\mu jj$, for same-sign charged leptons. This includes all leading logarithmic contributions which scale as
  $\alpha_W^4 \alpha_s^{2+k}\log^k(\hat s/p_\perp^2)$.  This process is
  important for the study of electroweak couplings and hence the QCD
  contributions are usually suppressed through a
  choice of Vector Boson Scattering (VBS) cuts. These select regions of phase space where logarithms in
  $\hat s/p_\perp^2$ are enhanced.
  While the logarithmic corrections lead to a small change for the
  cross sections, several
  distributions relevant for experimental studies are
  affected more significantly.
\end{abstract}

\newpage
\tableofcontents

\section{Introduction}
\label{sec:introduction}

\begin{figure}[bp]
  \centering
  \includegraphics[width=0.75\textwidth]{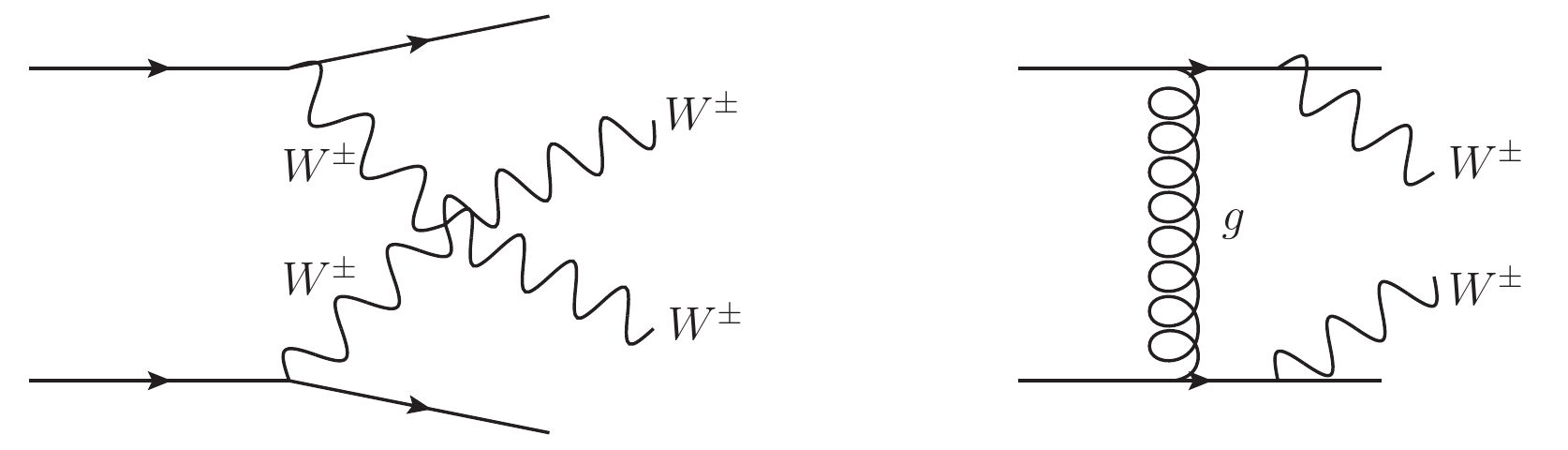}

  (a) \hspace{6.4cm} (b) \hspace{0.4cm} \phantom{a}
  \caption{Possible tree-level diagrams which contribute to $pp\to VVjj$ at (a)
    $\mathcal{O}(\alpha_W^4)$ and (b) $\mathcal{O}(\alpha_W^2 \alpha_s^2)$ to
    the squared matrix element.
    Interference between the two contributes at
    $\mathcal{O}(\alpha_W^3 \alpha_s)$.}
  \label{fig:treelevel}
\end{figure}

Vector boson scattering (VBS) describes the process $pp\to VV+2j$ where each $V$ may be
a $W$ or $Z$ boson.  All possible $V^*V^*\to VV$ processes can be inserted between two quark lines to
give this final state,
e.g.~figure~\ref{fig:treelevel}(a), and it provides a key opportunity to study the mechanism of
electroweak symmetry breaking, and in particular the means by which the Higgs
boson unitarises $VV$-scattering.  Other topologies may also contribute at this
order, $\mathcal{O}(\alpha_W^4)$ in the squared matrix element \emph{and} the
same final state may also be generated by gluon-exchange between
the quark lines which contributes at $\mathcal{O}(\alpha_W^2
\alpha_s^2)$.  There is also an interference channel between the two at $\mathcal{O}(\alpha_W^3
\alpha_s)$.  Given the experimental importance of this process, next-to-leading
order (NLO) QCD corrections to the different channels have been extensively
studied~\cite{Oleari:2003tc,Jager:2006zc,Jager:2006cp,Bozzi:2007ur,Jager:2009xx,Melia:2010bm,Denner:2012dz,Campanario:2013gea,Baglio:2014uba} and matched
with parton shower corrections~\cite{Melia:2011gk,Jager:2011ms,Jager:2013iza,Jager:2018cyo}.
Calculations of
NLO electroweak corrections~\cite{Biedermann:2016yds} and combined  NLO QCD and electroweak corrections are now also
available~\cite{Biedermann:2017bss,Denner:2019tmn,Chiesa:2019ulk,Denner:2020zit,Denner:2021hsa}.  See
e.g.~\cite{Covarelli:2021gyz} for a recent review.

In order to cleanly study the electroweak (EW) processes of interest, analyses typically apply cuts
which require a large invariant mass between two jets and/or a large rapidity
separation between two
jets~\cite{Aaboud:2019nmv,Sirunyan:2020gyx,Sirunyan:2020gvn}.  These are very effective
at enhancing the EW contributions over the QCD channels, but it is important to
accurately predict the purity of the sample which remains.
In particular, the requirement of large invariant mass or large rapidity
separation enhances the importance of logarithmic corrections of the form
$\log(\hat s/p_t^2)$ which arise at all orders of $\alpha_s$ (the leading
logarithmic (LL) corrections scale as
$\alpha_s^{k+2}\log^k(\hat s/p_t^2)$).  The combination $\alpha_s \log(\hat s/p_t^2)$ is
not necessarily small in these regions of phase space so that the logs can
damage the convergence of the perturbative expansion.  This means that the very
cuts which suppress the QCD contribution make it harder to calculate a reliable
estimate of that contribution.
%A community-wide study of predictions of the QCD
%component of VBS found large variations between different parton shower
%descriptions~\cite{Ballestrero:2018anz}, which may be a sign of perturbative
%instability.

Similar cuts are applied to separate the electroweak vector boson fusion (VBF) and QCD gluon fusion (GF) components in vector
boson fusion analyses of $pp\to H+2j$.  In this case, predictions including all
LL corrections and matched to fixed order are available within the \HEJ
framework~\cite{Andersen:2009nu,Andersen:2009he,Andersen:2017kfc,Andersen:2018tnm}.
Further, that framework also allows the inclusion of finite quark-mass effects for
arbitrary jet multiplicities~\cite{Andersen:2018kjg} which is currently only
available at LO up to 3-jets.  Here it was seen that while the inclusive cross
sections for GF were in good agreement between the LL predictions of
\HEJ and fixed-order calculations at NLO, significant differences in the invariant-mass and
rapidity distributions lead to a large difference in the predicted cross sections after cuts
on invariant-mass and rapidity-separation were applied. At NLO, the
impact of requiring  $|\Delta y_{j_1j_2}|>2.8$ and $m_{j_1j_2}>400$~GeV reduced the cross section to 9\% of the inclusive value
while at LL with finite quark masses, the impact of the same cuts reduced the
cross section to 4\% of the corresponding inclusive value, indicating the cuts were roughly a factor of two \emph{more}
effective. This behaviour is largely due to the dominance of the
$gg$-component within the inclusive cuts, and the difference in the
$m_{jj}$-spectrum of the $gg$, $qg$ and $qq$ components\cite{DelDuca:2001eu,DelDuca:2001fn}.
The same behaviour has also been seen in other comparisons
between combinations of fixed-order, parton shower and \HEJ predictions,
e.g.~\cite{Andersen:2014efa,Badger:2016bpw,Bendavid:2018nar}.

Given the identical type of cuts usually applied to the jets when studying VBS to those
in VBF, one would expect the impact of adding LL corrections to also be significant
in this process.
In this paper, we calculate the
$\alpha_s^{k}\log^k(\hat s/p_t^2)$ corrections to the
$\mathcal{O}(\alpha_W^2\alpha_s^2)$ component of VBS.  We do this within the
\HEJ framework.  These could then be straight-forwardly combined with
predictions for electroweak channels to give a complete description of
$pp\to W^\pm W^\pm +\ge2j$\footnote{We use this notation from now on to
  represent \emph{same-sign} $WW$ production in association with jets, i.e.~it
  is shorthand for $pp\to VV+\ge2j$, $V=W^+$ or $V=W^-$.}, for example using the method described
in~\cite{Chiesa:2019ulk}.  This was done previously for $pp\to W+2j$
in~\cite{Aaboud:2017fye} where LL predictions of the QCD component from \HEJ were combined with
electroweak predictions from
\textsc{Powheg+Pythia}~\cite{Nason:2004rx,Frixione:2007vw,Campbell:2013vha,Schissler:2013nga}
and were found to describe data very well, especially at large invariant mass.

In the following section, we outline how LL accuracy is obtained in the
predictions from \HEJ and then focus on how to construct the necessary
amplitudes to describe $pp\to W^\pm W^\pm+\ge 2j$ at this order.  In
section~\ref{sec:impact-ll-corr}, we illustrate the impact of the LL corrections
on distributions which are typically studied at the LHC,
both before and after VBS cuts are applied. We conclude in section~\ref{sec:conclusions}.

%%% Local Variables:
%%% mode: latex
%%% TeX-master: "main"
%%% End:

\section{Construction of Leading-Logarithm Amplitudes}
\label{sec:constr-ll-ampl}

\subsection{Logarithmic Accuracy}
\label{sec:logarithmic-accuracy}
%% identify the leading s/t for the born level process.
The Born-level scattering amplitudes for the same-sign $W$-pair production process
${pp\to W^\pm W^\pm+\ge2j}$ depicted in figure~\ref{fig:treelevel}(b) can be
expanded in powers of $\hat{s}/p_t^2$ considered in the limit of large $\hat{s}\to\infty$
with fixed $p_{\perp j_1}\sim p_{\perp j_2}\sim p_\perp$, where $p_\perp$ is
an arbitrary (but fixed) transverse scale. In this limit, large $\hat{s}$ with
fixed $p_\perp$ is reached by increasing rapidity differences between the
jets. For processes permitting a gluon (colour-octet) exchange between the
two jets (which the production of a same-sign $W$-pair does), this expansion of
$|\mathcal{M}|^2$ starts at $\alpha_s^2 \hat{s}^2/p_\perp^4$.

The leading logarithmic corrections to the cross section are controlled by the
scattering amplitudes in this so-called Multi-Regge Kinematic limit, where the
invariant mass between each pair of coloured particles is large but the
transverse scales are fixed. The higher-order corrections to the cross section
for these processes will contain terms of
$\alpha_s\log(\hat{s}/p_t^2)$\cite{Fadin:1975cb,Kuraev:1976ge,Kuraev:1977fs}, and
such corrections have traditionally been resummed through the use of the
formalism developed by Balitskii, Fadin, Kuraev and Lipatov\cite{Balitsky:1978ic}.

The all-order, leading logarithmic accuracy in $\log(\hat s/p_t^2)$ is
ensured by a systematic control of the power-expansion to
$\hat s/(p_t^{2})^n$ of $|\mathcal{M}|^2$ of each real correction with $n$
legs, and the expansion into terms of $(\alpha_s \log(\hat s/p_t))^n$ of the
virtual corrections, with subtraction terms organising the cancellation of
divergences between real and virtual corrections. In order to control just
the leading logarithmic corrections to the Born-level process, one needs to
consider just the leading expansion in $\hat{s}/p_t^2$ of each multiplicity, and
its virtual corrections. This leading term in the expansion in $\hat{s}/p_t^2$ has
contributions only from processes which allow a colour octet exchange
between each particle considered in order of increasing rapidity. So
processes with the rapidity ordering (or equivalently light-cone momentum
ordering in the MRK limit) $qQ\to Qq$ (incoming and outgoing states ordered
in light-cone momentum, and dropping the $W$s from the listing, since their
ordering is irrelevant for the discussion) will not contribute at leading
order in the expansion, whereas the leading order term \emph{will} receive
contributions from the same partonic process with the states ordered as
$qQ\to qQ$.

For states of higher multiplicity, the leading contribution is from
orderings such as $qQ\to qgQ$, which permit a colour octet (gluon, spin-1)
exchange between each parton considered in the ordering of rapidity. The
orderings $qQ\to gqQ$ require a colour \emph{singlet} (spin-$\frac 1 2$)
exchange between the $gq$. The contribution to the square of the scattering
amplitudes in the MRK limit is then suppressed by one power of $\hat{s}$ compared
to the contribution from the ordering $qQ\to qgQ$. This power suppression of
the amplitude leads to a logarithmic suppression of the contribution to the
cross section (see
e.g.~\cite{DelDuca:1995zy,Andersen:2017kfc,Andersen:2020yax}).

The factorisation of the scattering amplitudes in the MRK
limit\cite{Kuraev:1976ge,DelDuca:1995zy} implies that the MRK limit of these
amplitudes can be constructed to the required accuracy in $s/(p_t^2)^n$ using
building blocks which depend on a reduced set of the momenta. The cross
section can be calculated to leading logarithmic accuracy using an amplitude
consisting of so-called \emph{impact factors}, describing the forward and
backward particle production, and a \emph{Lipatov vertex} describing the
production of a gluon of central rapidity. The Lipatov vertex depends on the
momenta of the incoming and outgoing quarks, and the emitted gluon. The
impact factor\footnote{In the formalism of HEJ this impact factor is a
  current.} depends on the momenta of the incoming quark $p_a$, the outgoing
quark $p_1$ and the $W$ (or its decay products) only.  These impact factors
are in fact the same as those used for single $W$ production in association with
dijets\cite{Andersen:2009nu,Andersen:2012gk}. The cross sections discussed
in the current study will be matched to full fixed-order accuracy of high
multiplicity (up to 6 jets), see section~\ref{sec:matching-fixed-order}. Earlier studies have shown that with such
high-multiplicity matching, the overall impact of sub-leading logarithmic
contributions on two and three-jet observables is very minor
indeed. Therefore we will in this study calculate to just leading logarithmic
accuracy, matched to high multiplicity fixed order.

\boldmath
\subsection{Amplitudes for $pp\to W^\pm W^\pm+\ge2j$ Within HEJ}
\label{sec:amplitudes-ppto-wpm}
\unboldmath

In this subsection we present the new amplitudes required to apply the \HEJ
method to the process $pp\to W^\pm W^\pm+\ge2j$.  At this point we include the
leptonic decay of the $W$ bosons as we will need this to compute predictions for
typical experimental cuts and analyses.  As described in the previous
subsection, for a given multiplicity, a \HEJ amplitude is built out of impact
factors and Lipatov vertices.  These multiplicities are then combined with the
Lipatov ansatz for virtual corrections in order to achieve leading logarithmic
accuracy in $\hat{s}/p_t^2$ at all orders in \as.

We begin with the impact factors
which are independent of the number of gluons in the amplitude and hence can be
derived from the lowest order process where they occur.  The starting point
is therefore the leading order (LO) process:
\begin{align}
  \label{eq:momww}
  q(p_a)Q(p_b) \to (W_1^\pm\to) \ell(p_{\ell_1}) \bar\ell(p_{\bar\ell_1})
  (W_2^\pm\to) \ell'(p_{\ell_2}) \bar\ell'(p_{\bar\ell_2}) q'(p_1) Q'(p_2),
\end{align}
where $q$ and $Q$ represent different quark or anti-quark flavours.  There are
eight Feynman diagrams which contribute at LO, each similar to figure~\ref{fig:treelevel}(b), which arise from the $qq'W$ vertices for each boson being assigned to different points on different quark lines.

\begin{figure}
  \centering
  \includegraphics[width=0.65\textwidth]{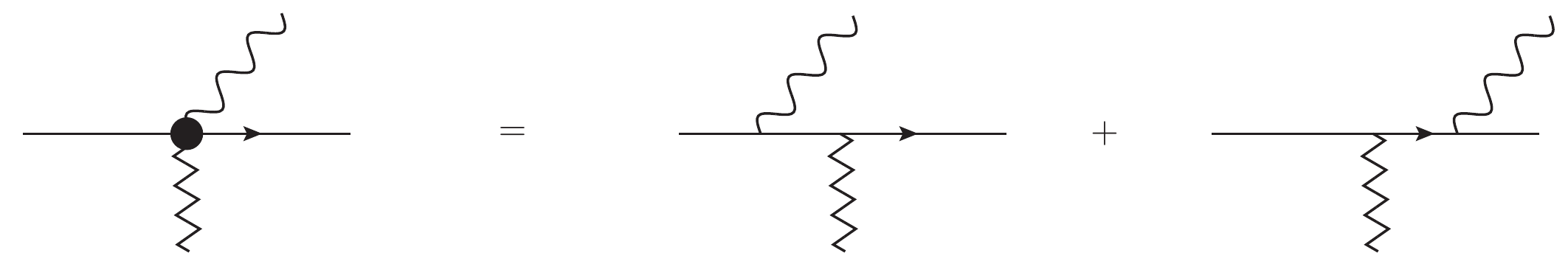}
  \caption{Schematic illustration of the current, $j^W_\mu(p_i,p_o,p_\ell,p_{\bar\ell})$, defined in Eq.~\eqref{eq:jW}
    to describe the production of a $W$ boson from a quark line with an
    off-shell gluon (shown as a zigzag line).}
  \label{fig:jW}
\end{figure}

We define the following current to describe the production of a $W$ boson from a quark
line with an off-shell gluon, $q(p_i)\to (W\to \ell \bar\ell) q'(p_o) g^*$:
\begin{align}
  \label{eq:jW}
  \begin{split}
    j^W_\mu(p_i,p_o,p_\ell,p_{\bar\ell}) =&\ \frac{g_W^2}{2}\ \frac{1}{(p_\ell +
      p_{\bar\ell})^2-m_W^2+i \Gamma_W m_W}\ \left[ \bar u^-(p_\ell) \gamma_\alpha
      v^-(p_{\bar\ell}) \right] \\
    &\times  \left( \frac{ \bar u^-(p_o) \gamma^\alpha
        (\slashed{p}_o+\slashed{p}_\ell + \slashed{p}_{\bar\ell}) \gamma_\mu
        u^-(p_i)}{(p_o + p_\ell + p_{\bar\ell})^2} +\frac{ \bar u^-(p_o) \gamma_\mu
        (\slashed{p}_i -\slashed{p}_\ell - \slashed{p}_{\bar\ell}) \gamma^\alpha
        u^-(p_i)}{(p_i - p_\ell - p_{\bar\ell})^2} \right),
  \end{split}
\end{align}
as illustrated in figure~\ref{fig:jW}.
The exact tree-level result can then be compactly expressed as the following two
contractions of two such currents:
\begin{align}
  \begin{split}
    \label{eq:HEJBorn}
    i\mathcal{M}^{\rm HEJ,tree}= g_s^2 &\left( \frac{j_\mu^W(p_a,p_1,p_{\ell_1},p_{\bar\ell_1})
        g^{\mu\nu} j_\nu^W(p_b,p_2,p_{\ell_2},p_{\bar\ell_2})}{q^2} \right. \\
    & \left. \qquad \qquad +\ \frac{j_\mu^W(p_a,p_1,p_{\ell_2},p_{\bar\ell_2})
        g^{\mu\nu} j_\nu^W(p_b,p_2,p_{\ell_1},p_{\bar\ell_1})}{\tilde q^2} \right),
  \end{split}
\end{align}
where $q=p_a-p_1-p_{\ell_1}-p_{\bar\ell_1}$ and
$\tilde q = p_a-p_1-p_{\ell_2}-p_{\bar\ell_2}$.
This amplitude remains exact at $\mathcal{O}(\alpha_W^2 \alpha_s)$ within the
\HEJ framework.  In particular, our
full amplitudes already achieve LO accuracy without the need for further
matching with the exception of channels with identical leptons or quarks.
The extra contributions arising in these special cases are suppressed in
the MRK limit and do not affect the logarithmic accuracy.  They are
nonetheless included through the fixed-order matching described in the next subsection.

In order to achieve LL accuracy in the inclusive predictions, we supplement the
tree-level amplitude
with corrections arising from real and virtual corrections at $\as^3$ and above.  Although separately
divergent in $4d$, we follow the procedure outlined
in~\cite{Andersen:2009nu,Andersen:2016vkp} to arrive at a finite amplitude for
$qQ\to W^\pm W^\pm q' g...gQ'$ for any number of intermediate gluons.  It is
built from~\cite{Andersen:2009nu}:
\begin{itemize}
\item a skeleton function for the process (often the LO matrix-element), given
  by a contraction of impact factors
\item a Lipatov vertex, $V^\mu$, for each real gluon emission with transverse momentum
  above scale $\lambda_{cut}$,
\item finite exponential factors for each $t$-channel propagator which arise from the sum of
  the virtual corrections at LL (given by the Lipatov Ansatz) and integration
  over all unresolved real emissions with $p_t<\lambda_{cut}$.
\end{itemize}
We typically choose values of $\lambda_{cut}$ to be around 0.2~GeV,
and have checked that our final results (cross sections and distributions) are
not strongly dependent on this value.  We have
  compared results for the total cross section at values of $\lambda_{cut}$
  between 0.2~GeV and 2~GeV and observe discrepancies of a few percent at most,
  with no clear trend.  This is a non-trivial check that the
implementation is correct.

The construction above is built around the concept of effective $t$-channel
momentum (although far more than contributions from $t$-channel diagrams are
included).  These are the momenta which would correspond to a planar $t$-channel
diagram when the outgoing coloured particles are ordered in rapidity.  For
$qQ\to W^\pm W^\pm q' g...gQ'$, it is already clear at lowest order that there
is not a unique definition of a $t$-channel momentum, as it depends on the
pairing of $W$ bosons with quark lines (see Eq.~\eqref{eq:HEJBorn}).  However,
the addition of extra gluons does not make this problem any worse than at
leading order, and hence for any number of coloured particles in the final state
there will be two sets of planar $t$-channels.  We will keep both (and the
interference between them) which significantly complicates the expressions
compared to the \HEJ description of single-$W$
production~\cite{Andersen:2012gk}.  It is similar to the treatment of
$pp\to Z+2j$ in \HEJ~\cite{Andersen:2016vkp}.  Specifically, we write the
amplitude using two skeleton functions, each with its own tower of real and
virtual corrections.  Interference between these is immediately included upon
squaring the amplitude.

\begin{figure}[btp]
  \centering
  \includegraphics[width=0.49\linewidth]{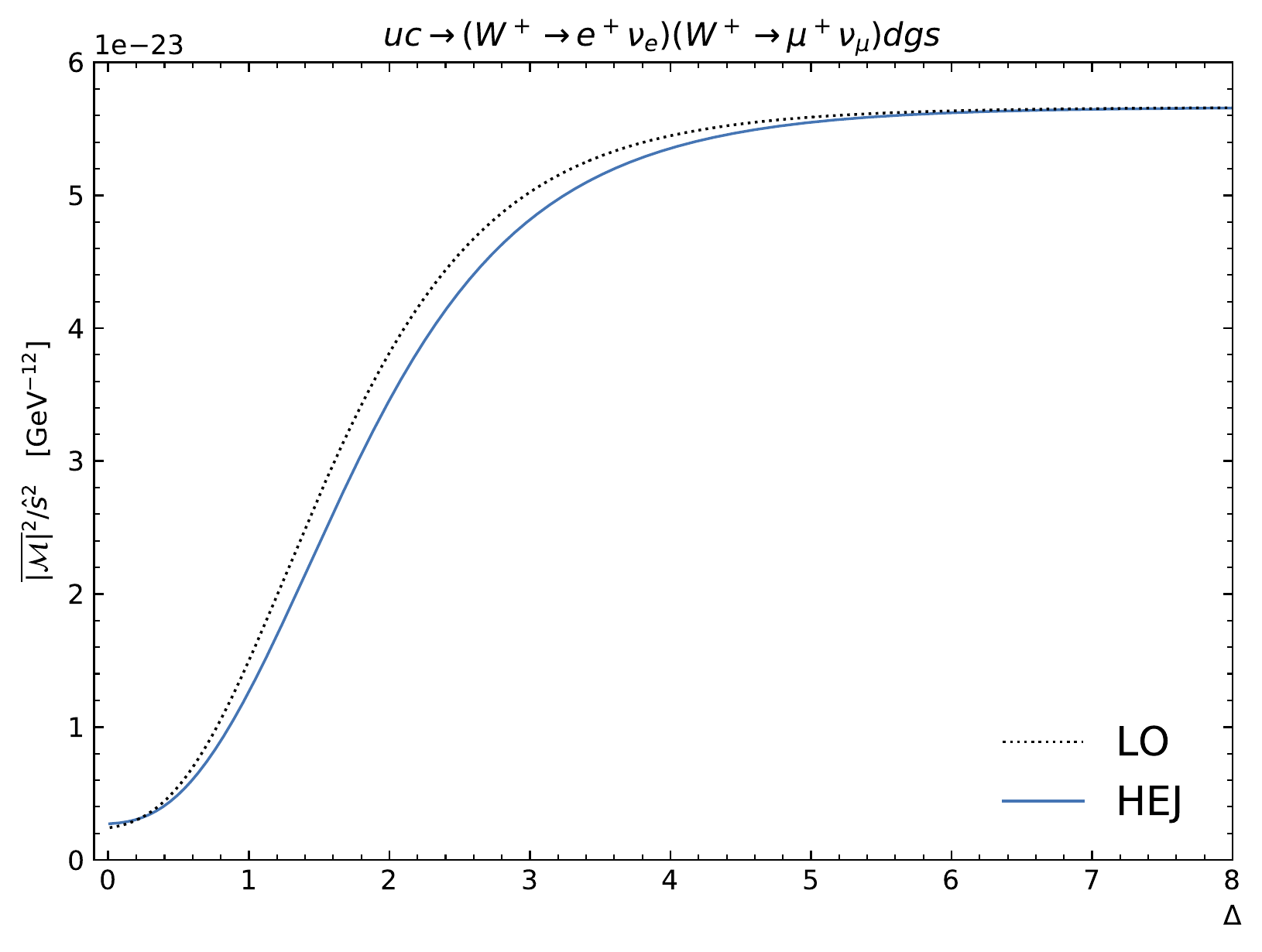}\hfill
  \includegraphics[width=0.49\linewidth]{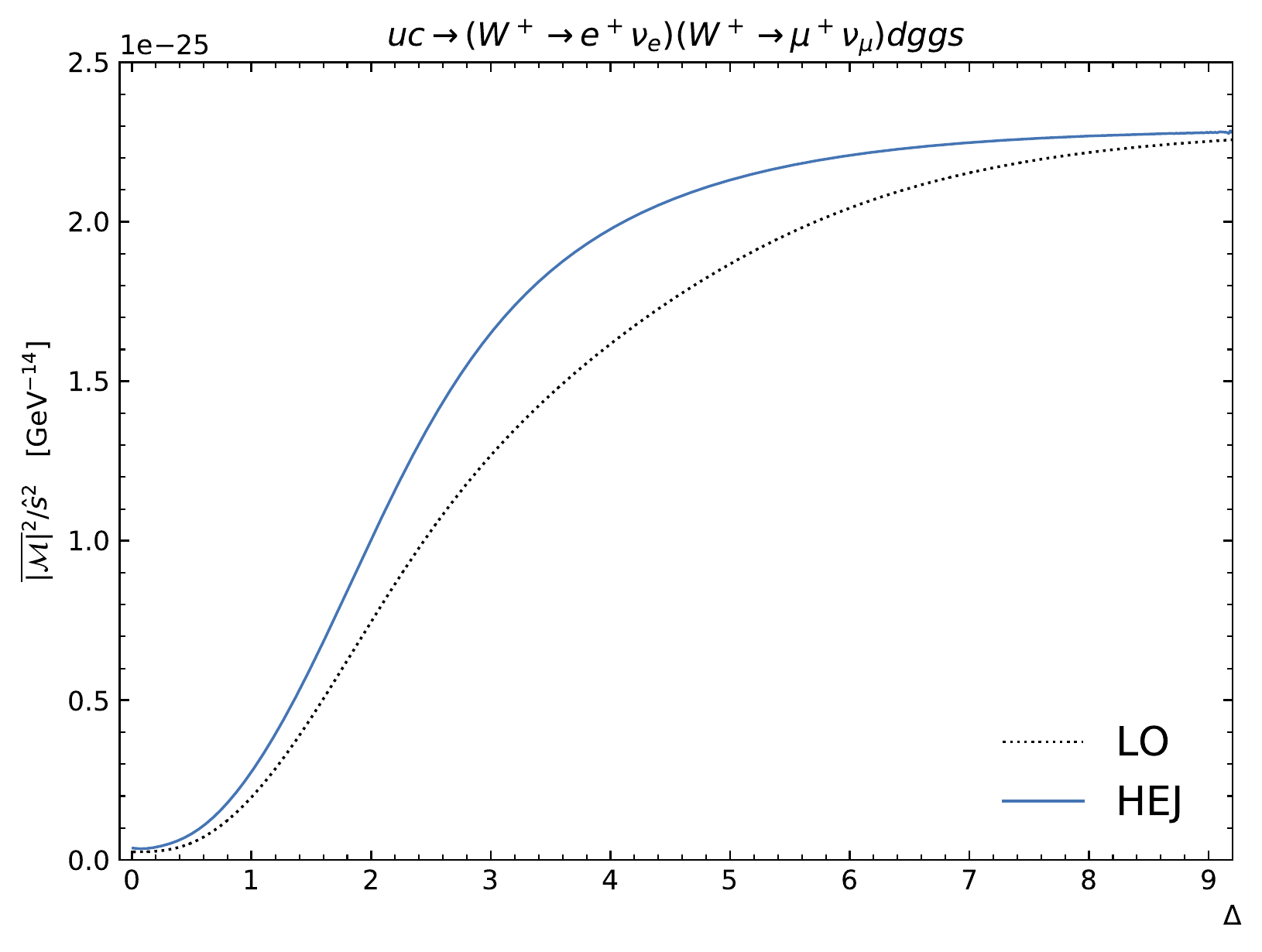}\\
  (a) \hspace{7cm} (b)
  \caption{Phase space explorer plots for $uc\to W^+ W^+ +n\textrm{ jets}$ for (a) 3 jets and (b) 4 jets.}
  \label{fig:explorers}
\end{figure}

These skeleton functions are defined as
\begin{align}
  \label{eq:Bdefintions}
  \begin{split}
    B &= j_\mu^W(p_a,p_1,p_{\ell_1},p_{\bar\ell_1}) g^{\mu\nu}
    j_\nu^W(p_b,p_n,p_{\ell_2},p_{\bar\ell_2}), \\
    \tilde B &= j_\mu^W(p_a,p_1,p_{\ell_2},p_{\bar\ell_2}) g^{\mu\nu}
    j_\nu^W(p_b,p_n,p_{\ell_1},p_{\bar\ell_1}),
  \end{split}
\end{align}
which relate to the two possible combinations of leptons and quark lines.  These
have corresponding planar $t$-channel momenta:
\begin{align}
  \label{eq:qs}
  \begin{split}
  q_1&=p_a-p_1-p_{\ell_1}-p_{\bar\ell_1}, \qquad q_i=q_{i-1}-p_i \quad
  i=2,...,n-1, \\
  \tilde q_1&=p_a-p_1-p_{\ell_2}-p_{\bar\ell_2}, \qquad \tilde q_i=\tilde q_{i-1}-p_i \quad
  i=2,...,n-1.
\end{split}
\end{align}
We refer to the corresponding momenta-squared as $t_i=q_i^2$ and $\tilde
t_i=\tilde q_i^2$ for $i=1,...,n-1$.  We also define the rapidity differences of consecutive quarks/gluons to be $\Delta y_i=y_{i+1}-y_i$.
The LL accurate matrix element is then given by
\begin{align}
  \label{eq:allordereg}
  \begin{split}
    &   |\mathcal{M}^{\rm HEJ,reg}_{qQ\to W_1^\pm W_2^\pm q'(n-2)gQ'}|^2\\ =&\ g_s^4 \frac{C_F}{8N_c}\ ( g_s^2
    \Ca)^{n-2}\  \\  &\times \Bigg( \frac{|B|^2}{t_{1}t_{(n-1)}}
    \exp(\omega^0(q_{(n-1)\perp})\Delta y_{n-1}) \prod^{n-2}_{i=1} \frac{-V^2(q_{i},
      q_{(i+1)})}{t_{i} t_{(i+1)}} \exp(\omega^0(q_{i\perp})\Delta y_i)\\
    &\quad +\  \frac{|\tilde B |^2}{\tilde t_{1}\tilde
      t_{(n-1)}} \exp(\omega^0(\tilde q_{(n-1)\perp})\Delta y_{n-1})
    \prod^{n-2}_{i=1}\frac{-V^2(\tilde q_{i}, \tilde q_{(i+1)})}{\tilde t_{i}
      \tilde t_{(i+1)}} \exp(\omega^0(\tilde q_{i\perp})\Delta y_i) \\
    &\quad+\ \frac{2\Re\{ B \tilde B\}}{\sqrt{t_{1}\tilde t_{1}}\sqrt{t_{(n-1)} \tilde
        t_{b(n-1)}}} \exp(\omega^0(\sqrt{q_{(n-1)\perp}\tilde q_{(n-1)\perp}})\Delta y_{n-1})\\
    & \hspace{3cm} \; \times\prod^{n-2}_{i=1}\frac{-V(q_{i}, q_{(i+1)})\cdot V(\tilde q_{i},
      \tilde q_{(i+1)})}{\sqrt{t_{i} \tilde t_{i}} \sqrt{t_{(i+1)}\tilde
        t_{(i+1)}}}\exp(\omega^0(\sqrt{q_{i\perp} \tilde q_{i\perp}})\Delta y_{i})\Bigg),
  \end{split}
\end{align}
where
\begin{align}
  \label{eq:omega0}
  \begin{split}
    V^\mu(q_1,q_2) &= -(q_1+q_2)^\mu +
    \frac{p_a^\mu}{2}\left(\frac{q_1^2}{p\cdot p_a} + \frac{p\cdot p_b}{p_a\cdot p_b}
      + \frac{p\cdot p_n}{p_a\cdot p_n} \right) + p_a \leftrightarrow p_1, \\
    & \qquad - \frac{p_b^\mu}{2} \left( \frac{q_2^2}{p\cdot p_b} + \frac{p\cdot
        p_a}{p_a\cdot p_b} + \frac{p\cdot p_1}{p_b\cdot p_1} \right) - p_b
    \leftrightarrow p_n  \qquad {\rm{for}}\ p=q_1-q_2, \\
  \omega^0(q_{\perp}) &= - \frac{g_s^2 \Ca}{4\pi^2} \log\left( \frac{q_\perp^2}{\lambda_{cut}^2}\right).
  \end{split}
\end{align}
One can immediately check that at
$\mathcal{O}(\alpha_W^4 \alpha_s^2)$ this exactly agrees with the summed, averaged and squared
amplitude of Eq.~\eqref{eq:HEJBorn}.  One can extend this test to higher orders
by comparing the result of
Eq.~\eqref{eq:allordereg} with the virtual corrections removed
(i.e.~setting $\omega^0(q_\perp^2)=0$) at a fixed order in
$\alpha_s$ to the corresponding LO result.  In the MRK limit, these should match.  We illustrate this in figure~\ref{fig:explorers} for squared matrix
elements for sample channels at $\alpha_W^4\alpha_s^3$ and $\alpha_W^4\alpha_s^4$.
The rapidities of the outgoing particles in these slices are:

\begin{tabular}[h]{cl}
  (a)\ & $y_d = y_e = y_{\nu_e} = \Delta, \ y_s = y_\mu = y_{\nu_\mu} = -\Delta,\ y_g = 0$, \\
  (b)\ & $y_d = y_e = y_{\nu_e} = \Delta, \ y_s = y_\mu = y_{\nu_\mu} = -\Delta,
         \ y_{g_1} = \Delta  / 3, \ y_{g_2} = - \Delta  / 3$. \\
\end{tabular} \\
The rapidity separation of the quarks and gluons is controlled by the
parameter $\Delta$ and hence the MRK limit is approached at the right-hand side of
the plots.   The other parameters used are given in Appendix~\ref{sec:moment-conf-phase}, but
the behaviour seen is not sensitive to the exact values of the transverse
momentum or the azimuthal angles.  In both plots, we show the squared matrix
element divided by $\hat s^2$, to achieve a finite non-zero value in the MRK
limit.  The exact LO result (black, dotted line) and the approximation within
\HEJ (solid, red line) are very similar throughout the range and converge to the
same limiting value at large $\Delta$.  The plots also show the rich dynamics of
the matrix elements which would be missed at small values of $\Delta$ if the
limiting value was used throughout phase space.

The LL accurate cross section is then given by the following sum over multiplicities and
integration over all phase space (where $\ell_l$ numbers the four leptons from
the $W$ decays):
\begin{align}
  \label{eq:sigmaWW}
  \begin{split}
    \sigma^{\rm LL}_{pp\to W^\pm W^\pm+2j} = &\sum_{f_{i1}, f_{i2}}\ \sum_{n=2}^\infty\
    \prod_{i=1}^n\left(\int \frac{\mathrm{d}^2\mathbf{p}_{i\perp}}{(2\pi)^3}\
      \int \frac{\mathrm{d} y_i}{2}
     \right)\ \prod_{l=1}^4\left(\int \frac{\mathrm{d}^2\mathbf{p}_{\ell_l
                                \perp}  }{(2\pi)^3}\
       \int \frac{\mathrm{d} y_{\ell_l}}{2}
    \right) \\
    & \times
    \frac{\overline{\left|\mathcal{M}^{\rm HEJ,reg}_{qQ\to W_1^\pm W_2^\pm q'(n-2)gQ'}\right|}^2}{\hat s^2} \\
    &\times\ \ x_a f_{A,f_{i1}}(x_a, Q_a)\ x_2 f_{B,f_{i2}}(x_b, Q_b)\ (2\pi)^4\ \delta^2\!\!\left(\sum_{k=1}^n
      \mathbf{p}_{k\perp} + \sum_{m=1}^4 \mathbf{p}_{\ell_m \perp}\right )\ \mathcal{O}_{2j}(\{p_j\}).
  \end{split}
\end{align}
We emphasise here that no approximation is made to the phase space being
integrated over, only within the matrix element itself.  This integral can be
efficiently implemented in an exclusive Monte Carlo event generator giving full
flexibility to implement experimental cuts and distributions.  Before this integration, we
first multiply the squared matrix element by reweighting factors to implement
fixed-order accuracy, as discussed in the next section.

\subsection{Matching to Fixed Order}
\label{sec:matching-fixed-order}

In the previous subsection, we have described how to construct the cross section
for $pp\to W^\pm W^\pm+\ge2j$ at LL accuracy in $\hat s/p_t^2$.  In order to
increase the validity of the approach, we will supplement this with subleading
terms which will provide leading-order accuracy at each order in $\alpha_s$, up
to the point where this is computationally feasible.  For this process, in this
study, that is samples with 2, 3, 4, 5 and 6 jets at LO.  We observe that the
impact of adding higher multiplicity fixed-order samples decreases with each
multiplicity, and in particular that the $6$-jet sample has at most a
few percent effect in any distribution so we are confident that our results have
converged.

The matching is then implemented using the methods of
\HEJtwo~\cite{Andersen:2018tnm} which reorganises the integral over phase space
to supplement fixed-order samples at each order with real and virtual
corrections such that leading-logarithmic accuracy is maintained at all orders
in $\alpha_s$ and additionally leading-order accuracy is achieved for the
$n$-jet components for $n=2$--$6$.  This method is implemented in the exclusive
event generator, \HEJtwo~\cite{Andersen:2019yzo}, which is publicly available at
\texttt{https://hej.hepforge.org}. The process $pp\to W^\pm W^\pm+\ge2j$ will be
included in a future public release of \HEJtwo. The fixed-order input is given
as Les Houches events and can be taken from any generator.  In this study we
have used Sherpa~\cite{Bothmann:2019yzt} to generate the fixed-order input.

Finally, we rescale all the final \HEJtwo predictions to match the total cross
section to the inclusive NLO cross section for each scale choice.  However, for
the setup described in section~\ref{sec:impact-ll-corr}, and as discussed there,
this turns out to have a negligible impact in this case.

%%% Local Variables:
%%% mode: latex
%%% TeX-master: "main"
%%% End:

\section{Impact of Leading-Logarithm Corrections}
\label{sec:impact-ll-corr}
We will now show the predictions constructed and matched as described in the
previous section for observables commonly studied at the LHC for the process
$pp\to W^\pm W^\pm +\ge2j$, where one $W^\pm$ decays in the electron channel and
one in the muon channel.  We will
compare these with an NLO calculation of the same process (here taken from
Sherpa~\cite{Bothmann:2019yzt} using COMIX~\cite{Gleisberg_2008} with the extension of
OpenLoops~\cite{Buccioni:2017yxi}) to assess the impact of the new LL corrections.
Also included for comparison is a calculation including MC@NLO matching~\cite{Frixione:2002ik} with the shower generator CS Shower~\cite{Schumann:2007mg} packaged with Sherpa. We
use the NNPDF3.0 NLO PDF set~\cite{Ball:2014uwa} as provided by LHAPDF6~\cite{Buckley:2014ana}. We choose the central factorisation and renormalisation scale as the geometric mean of the transverse momenta of the two leading jets, $\mu_F = \mu_R = \sqrt{p_{\perp;j_1}p_{\perp;j_2}}$. These scales are varied independently by a factor of two around this central value, with the constraint that their ratio is kept between 0.5 and 2. The uncertainty bands shown in the plots are obtained from the envelope of these variations.

\begin{figure}[btp]
\centering
 \begin{subfigure}{0.49\textwidth}
   \includegraphics[width=\textwidth]{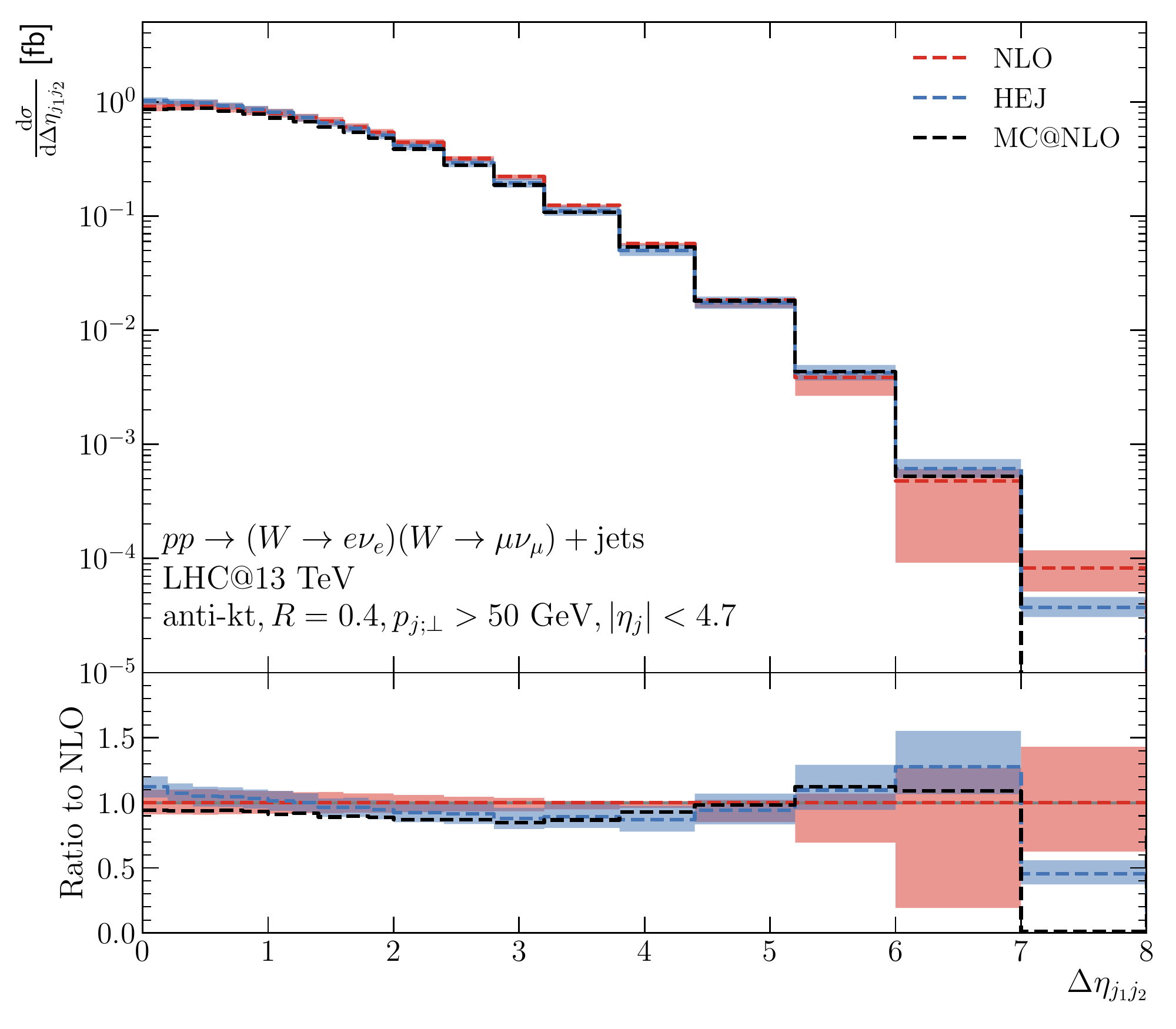}
   \caption{}
   \label{fig:deltay_jj_novbs}
 \end{subfigure}
 \begin{subfigure}{0.49\textwidth}
   \includegraphics[width=\textwidth]{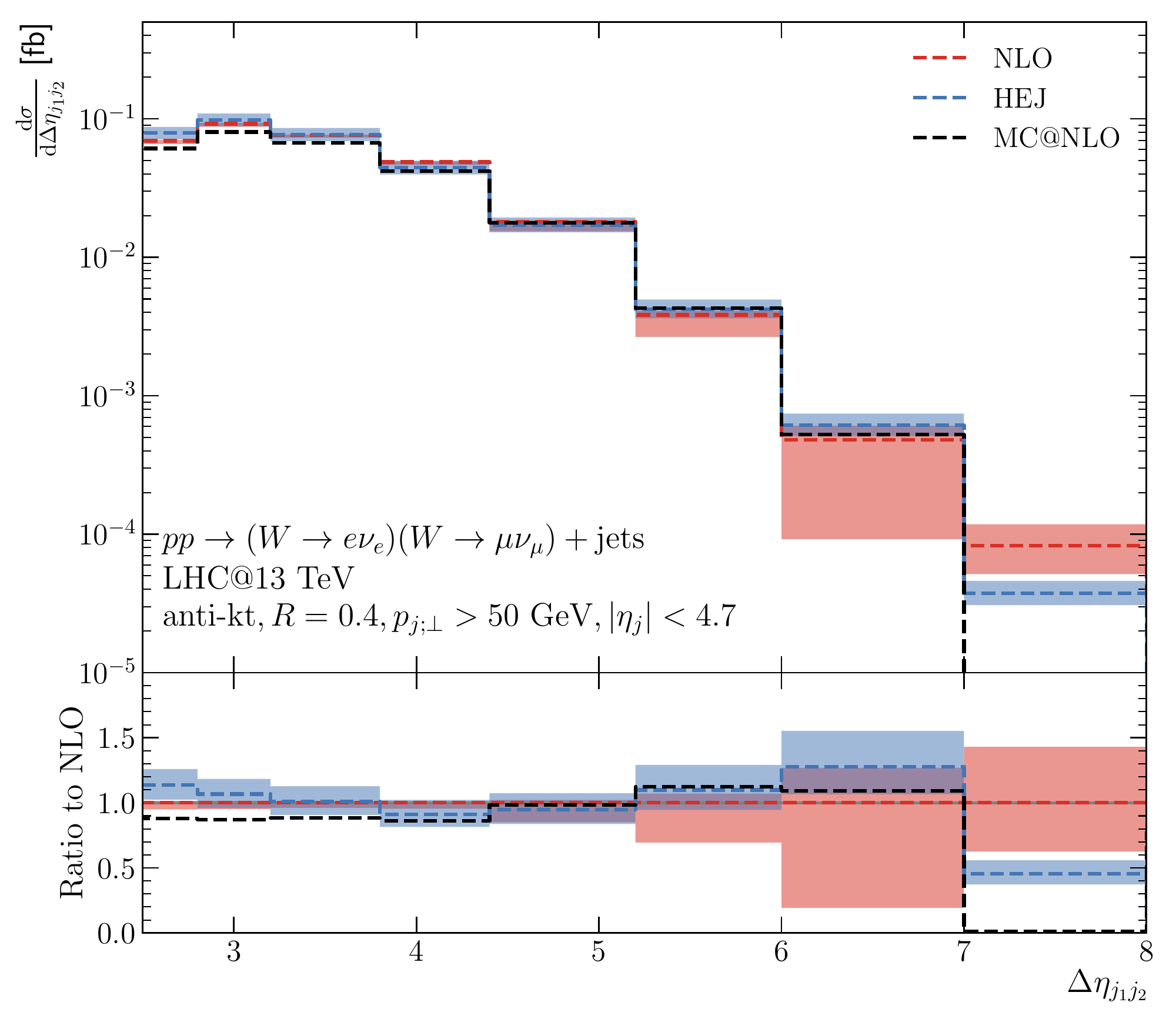}
   \caption{}
   \label{fig:deltay_jj_vbs}
 \end{subfigure}
 \caption{The differential distribution in the pseudorapidity separation of the two
leading jets in $pp\to W^\pm W^\pm +\ge2j$, (a) without and (b) with additional VBS cuts.}
 \label{fig:deltayjj}
\end{figure}

\begin{figure}[btp]
\centering
 \begin{subfigure}{0.49\textwidth}
   \includegraphics[width=\textwidth]{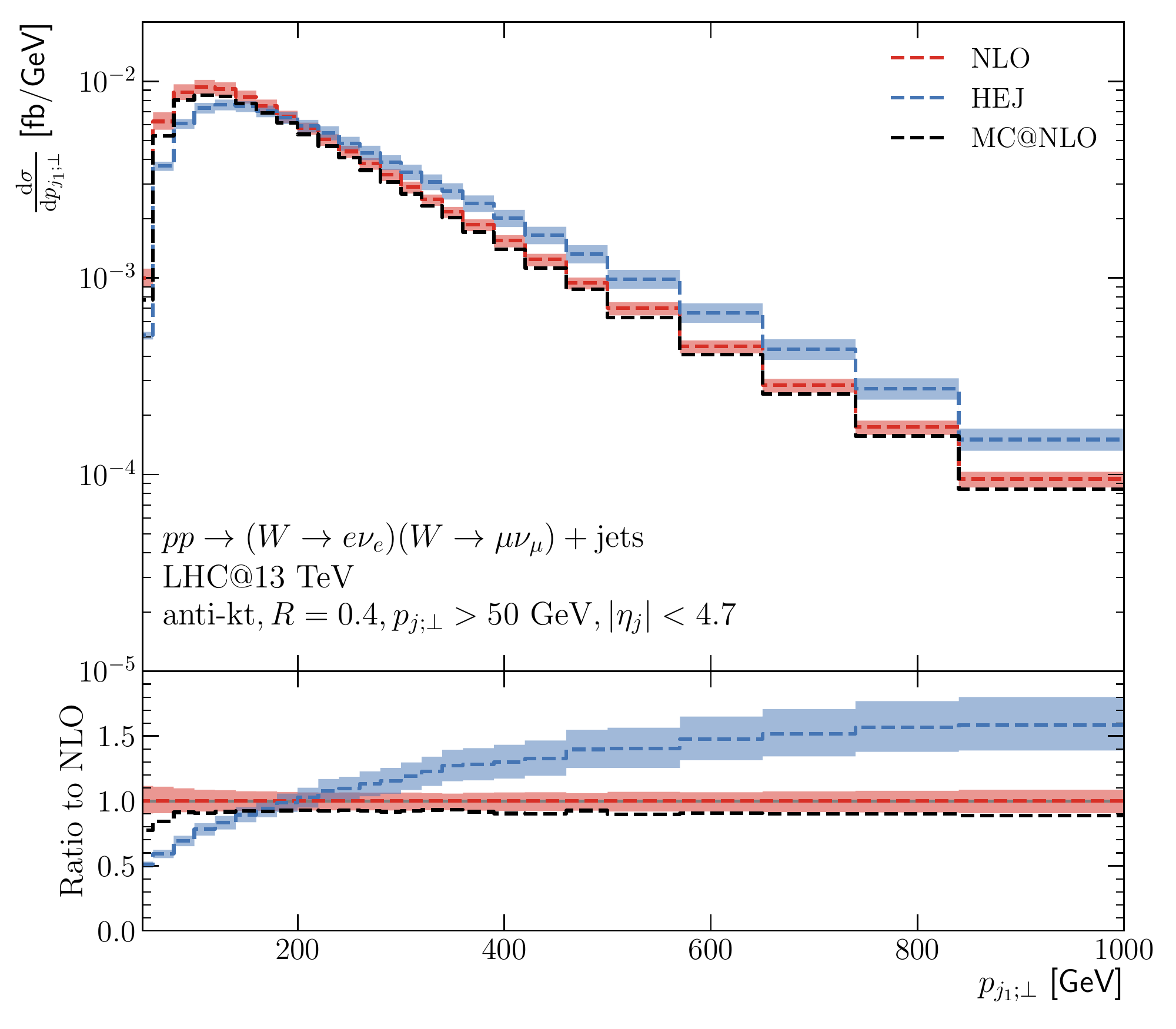}
   \caption{}
   \label{fig:jet1_pt_novbs}
 \end{subfigure}
 \begin{subfigure}{0.49\textwidth}
   \includegraphics[width=\textwidth]{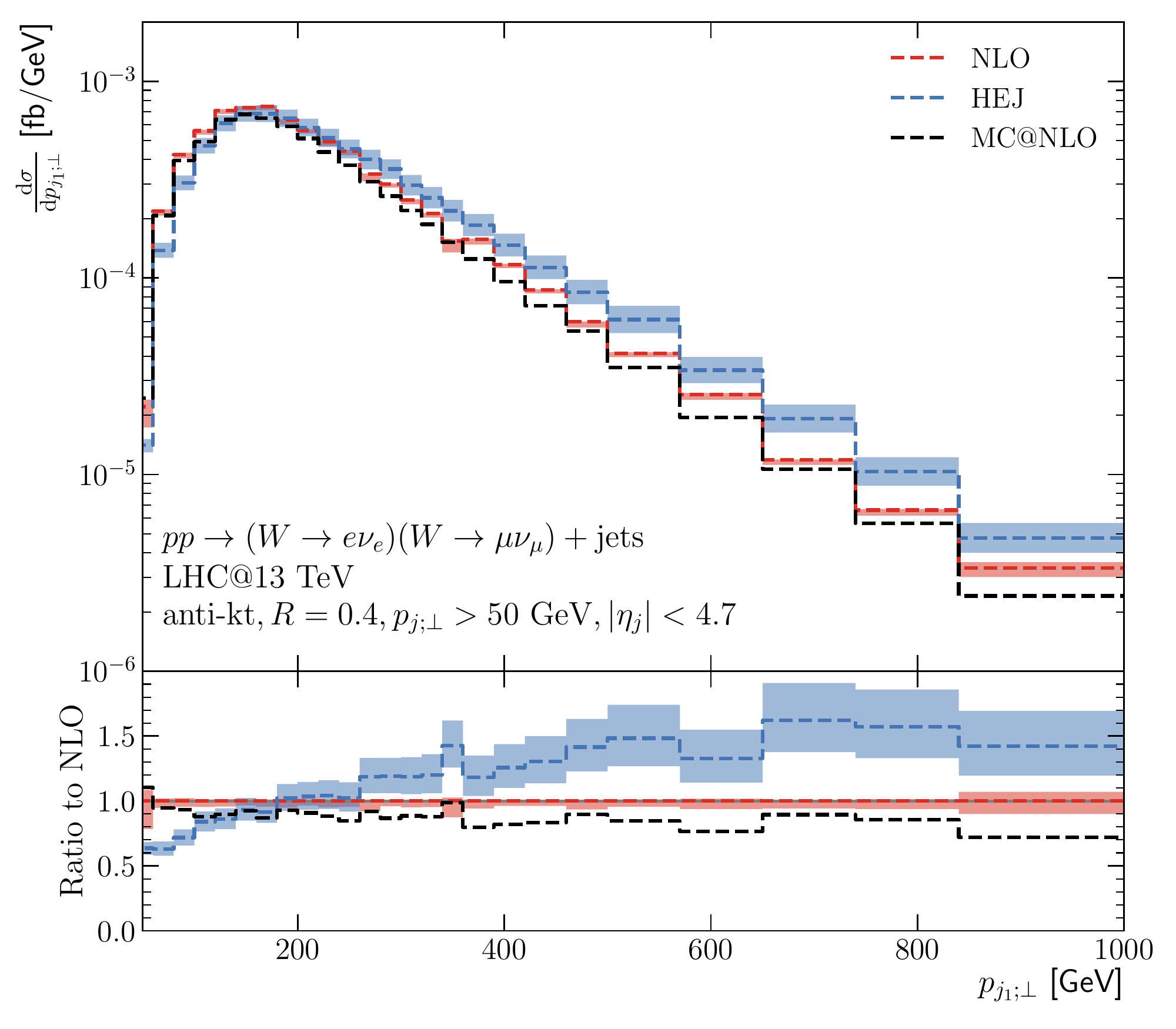}
   \caption{}
   \label{fig:jet1_pt_vbs}
 \end{subfigure}
 \caption{The differential distribution in the transverse momentum of the
   hardest jet in $pp\to W^\pm W^\pm +\ge2j$, (a) without and (b) with
   additional VBS cuts.}
 \label{fig:jet1_pt}
\end{figure}

\begin{figure}[btp]
\centering
 \begin{subfigure}{0.49\textwidth}
   \includegraphics[width=\textwidth]{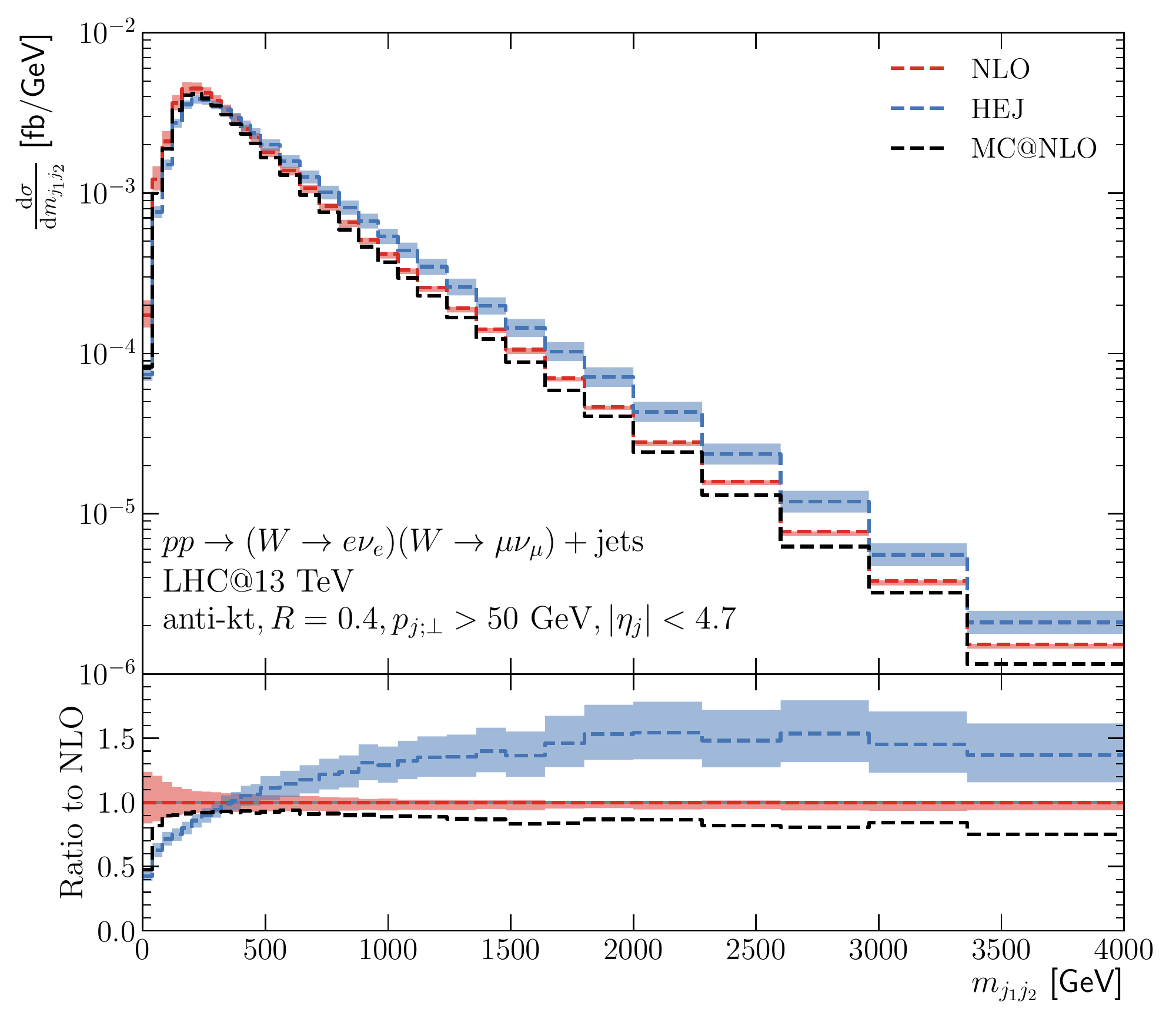}
   \caption{}
   \label{fig:m_j1j2_novbs}
 \end{subfigure}
 \begin{subfigure}{0.49\textwidth}
   \includegraphics[width=\textwidth]{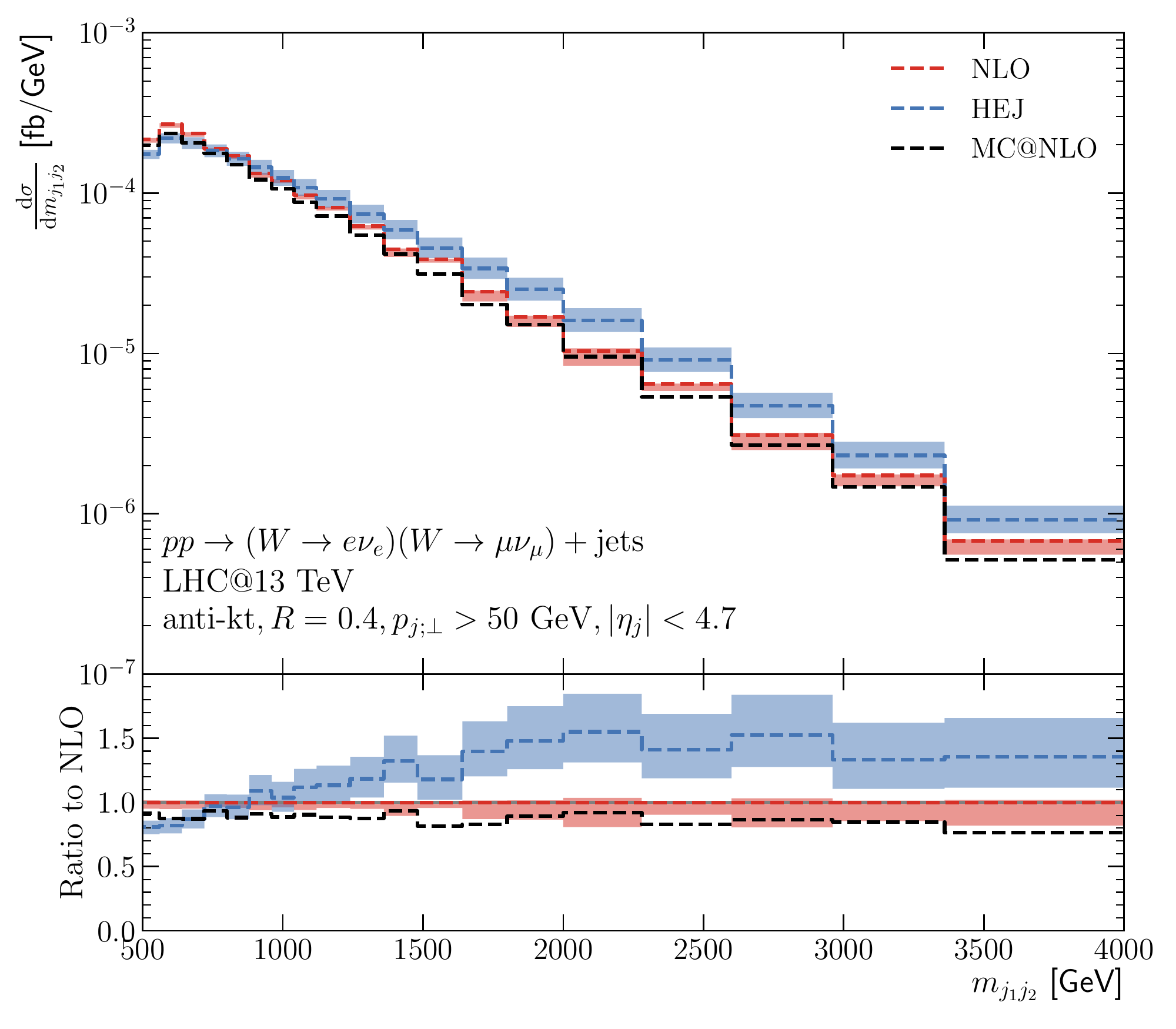}
   \caption{}
   \label{fig:m_j1j2_vbs}
 \end{subfigure}
 \caption{The differential distribution in the invariant mass of the two
leading jets in $pp\to W^\pm W^\pm +\ge2j$, (a) without and (b) with additional VBS
cuts.}
 \label{fig:m_j1j2}
\end{figure}

\begin{figure}[btp]
\centering
 \begin{subfigure}{0.49\textwidth}
   \includegraphics[width=\textwidth]{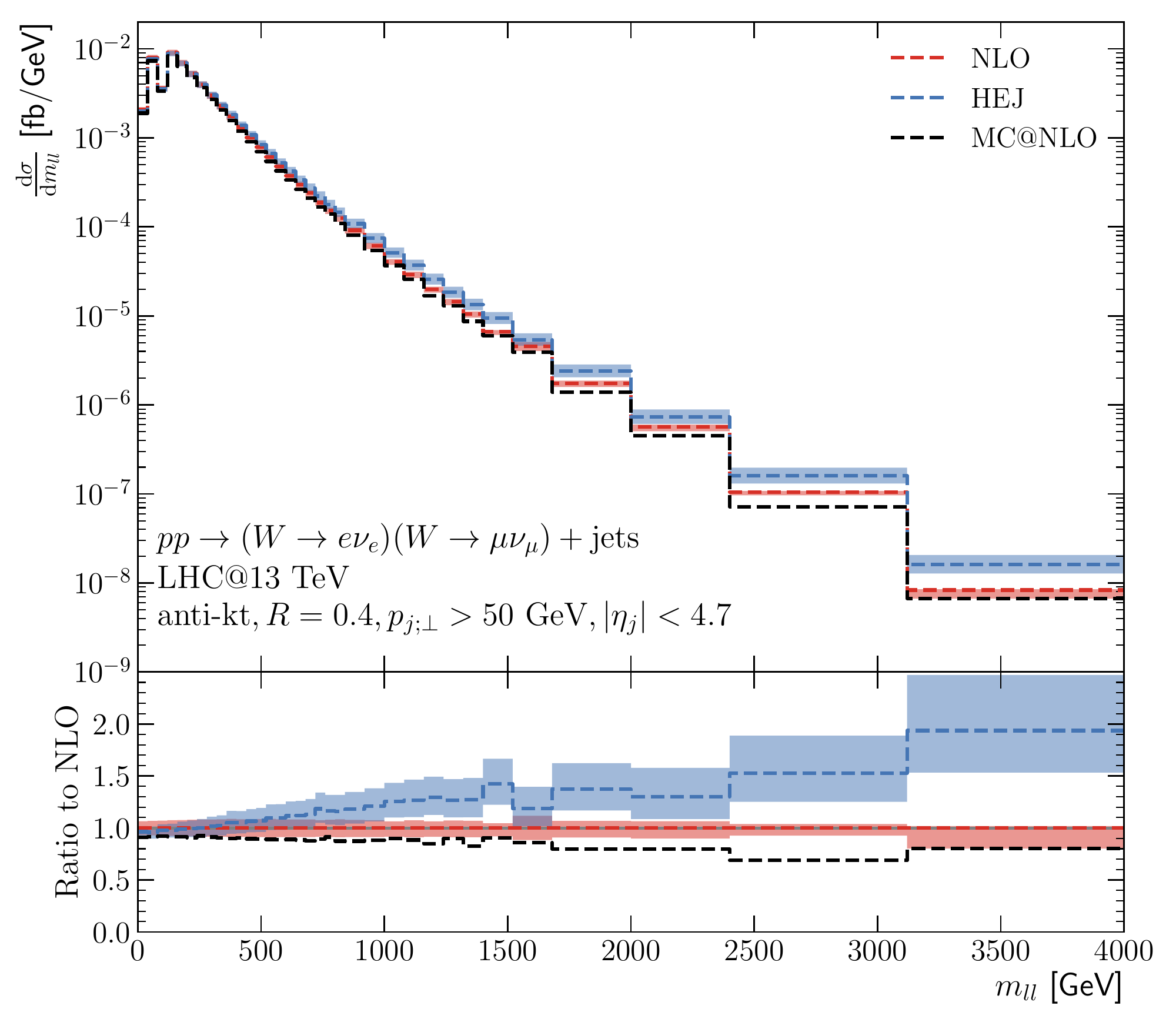}
   \caption{}
   \label{fig:m_l1l2_novbs}
 \end{subfigure}
 \begin{subfigure}{0.49\textwidth}
   \includegraphics[width=\textwidth]{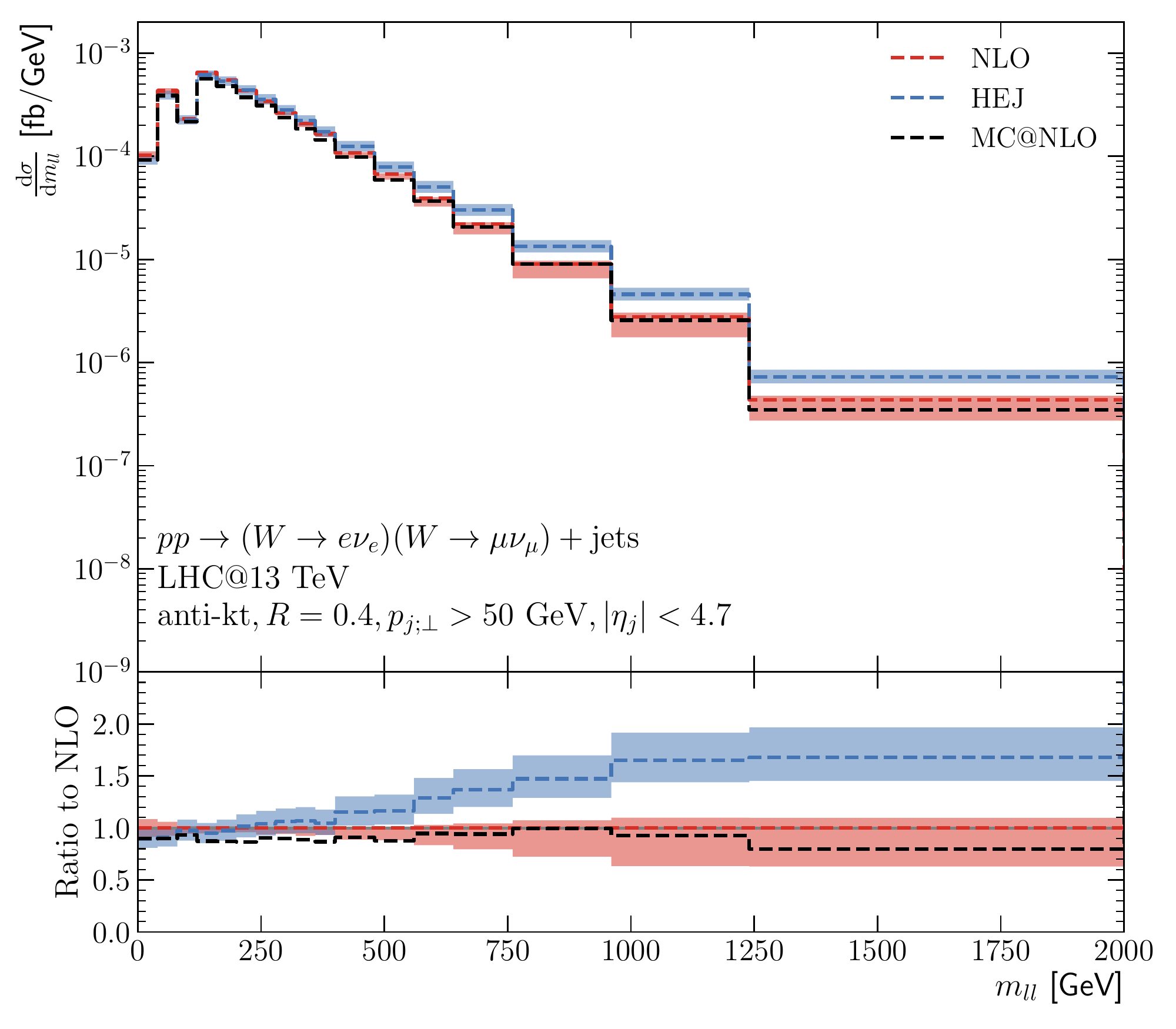}
   \caption{}
   \label{fig:m_l1l2_vbs}
 \end{subfigure}
 \caption{The differential distribution in the invariant mass of the two charged
leptons in $pp\to W^\pm W^\pm +\ge2j$, (a) without and (b) with additional VBS cuts.}
 \label{fig:m_l1l2}
\end{figure}

\begin{figure}[btp]
\centering
 \begin{subfigure}{0.49\textwidth}
   \includegraphics[width=\textwidth]{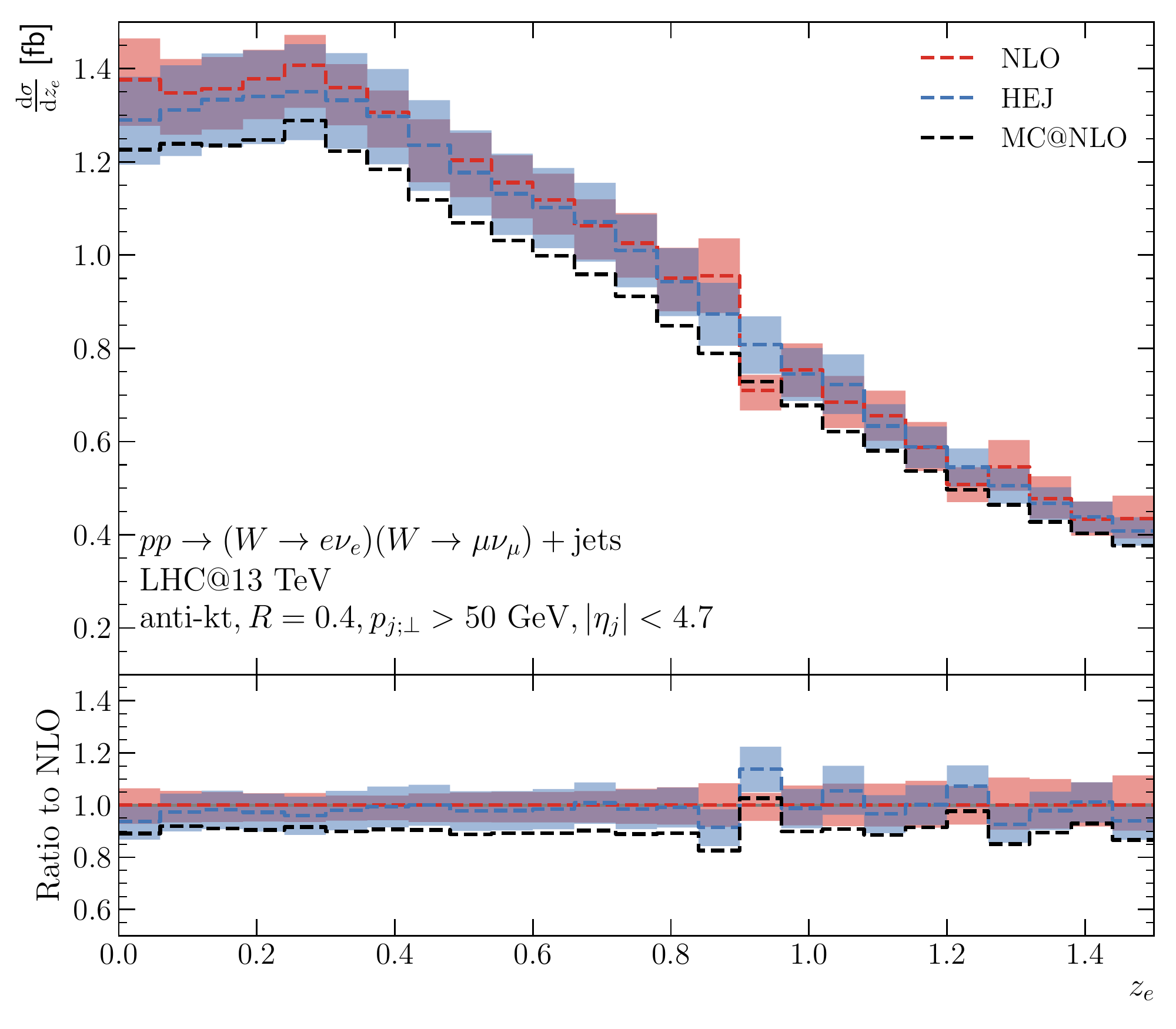}
   \caption{}
   \label{fig:zeppenfelde_novbs}
 \end{subfigure}
 \begin{subfigure}{0.49\textwidth}
   \includegraphics[width=\textwidth]{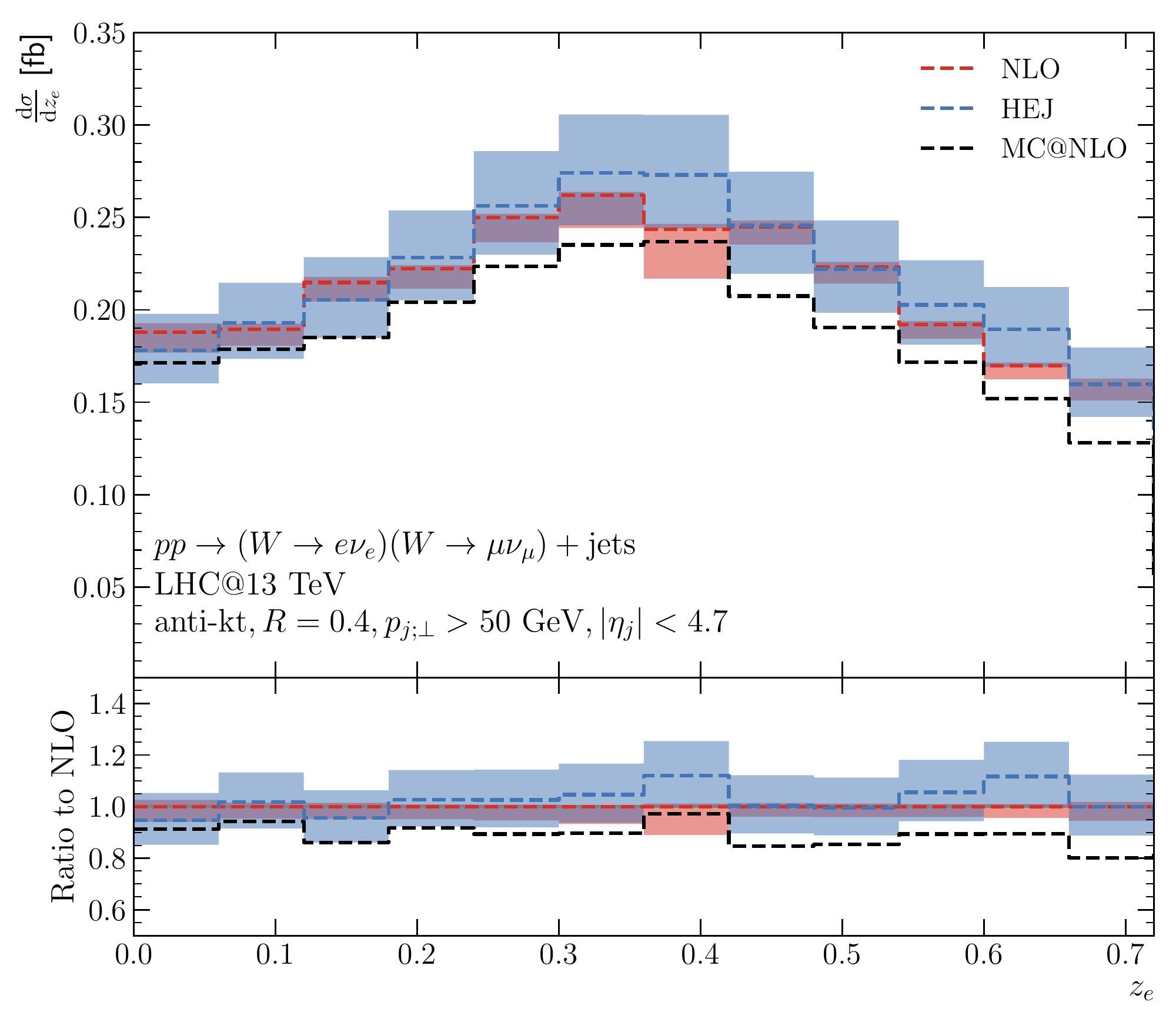}
   \caption{}
   \label{fig:zeppenfelde_vbs}
 \end{subfigure}
 \caption{The differential distribution in the Zeppenfeld variable for the electron in $pp\to W^\pm W^\pm +\ge2j$, without and with additional VBS cuts. }
 \label{fig:zeppenfelde}
\end{figure}

In figures~\ref{fig:deltayjj}--\ref{fig:zeppenfelde} we show distributions
measured in a recent CMS analysis~\cite{Sirunyan:2020gyx}.  It is not meaningful
to compare to the data points in that study as these include the
large $\mathcal{O}(\alpha_W^4)$ contributions.  The cuts applied to
the predictions are listed in Table~\ref{tab:cutsCMS}.  Those in the first group
form the inclusive cuts which are applied to all plots.  The additional three
criteria (below the second horizontal line) give the extra cuts on leading dijet
invariant mass, leading jet pseudorapidity separation and the Zeppenfeld variable~\cite{Rainwater:1996ud}
\begin{align}
  \label{eq:zepp}
  z_l = \frac{\eta_l -\frac12 (\eta_{j_1}+\eta_{j_2})}{|\eta_{j_1}-\eta_{j_2}|},
\end{align}
where $j_1$ and $j_2$ are the two hardest jets in the event. These cuts
are used to try to suppress the QCD contribution to this process.  We
will refer to them as ``VBS cuts'' and will show results before and
after these extra criteria.  The output of HEJ is exclusive in the momenta of all outgoing
particles.  Here we have used the functionality of linking HEJ directly with
Rivet\cite{Rivet:2013} to apply the cuts and fill the histograms.

\begin{table}[btp]
  \centering
  \begin{tabularx}{\linewidth}{l|l}
  \textbf{Variable} & \textbf{Selection Cut} \\
   \hline
    Lepton pseudorapidity & $|\eta_l|<2.5$\\
    Jet pseudorapidity & $|\eta_j|<4.7$\\
    Leading/subleading lepton $p_\textrm{T}$ & $p_T>25/20$~GeV \\
    Missing transverse momentum & $E^{\textrm{miss}}_{\textrm{T}}>30$~GeV\\
    Jet $p_\textrm{T}$& $p_\textrm{T}>50$~GeV\\
    Lepton isolation &$\Delta R(l,\textrm{jet})>0.4$ o/w jet is removed\\
    Di-lepton mass & $m_{ll}>20$~GeV \\
    Di-lepton mass restriction & $|m_{ll}-m_{Z}|>15$~GeV \\
    \hline
    Di-jet mass & $m_{j_1j_2}> 500$~GeV \\
    Jet rapidity separation & $|\Delta \eta_{j_1j_2}|> 2.5$\\
    Max lepton Zeppenfeld variable (Eq.~\ref{eq:zepp}) & $\max(z_l) < 0.75$
  \end{tabularx}
  \caption{The selection cuts used in the analysis where the lepton cuts apply only to the charged leptons. The last three rows define the additional VBS cuts.}
  \label{tab:cutsCMS}
\end{table}

\begin{figure}[btp]
\centering
 \begin{subfigure}{0.49\textwidth}
   \includegraphics[width=\textwidth]{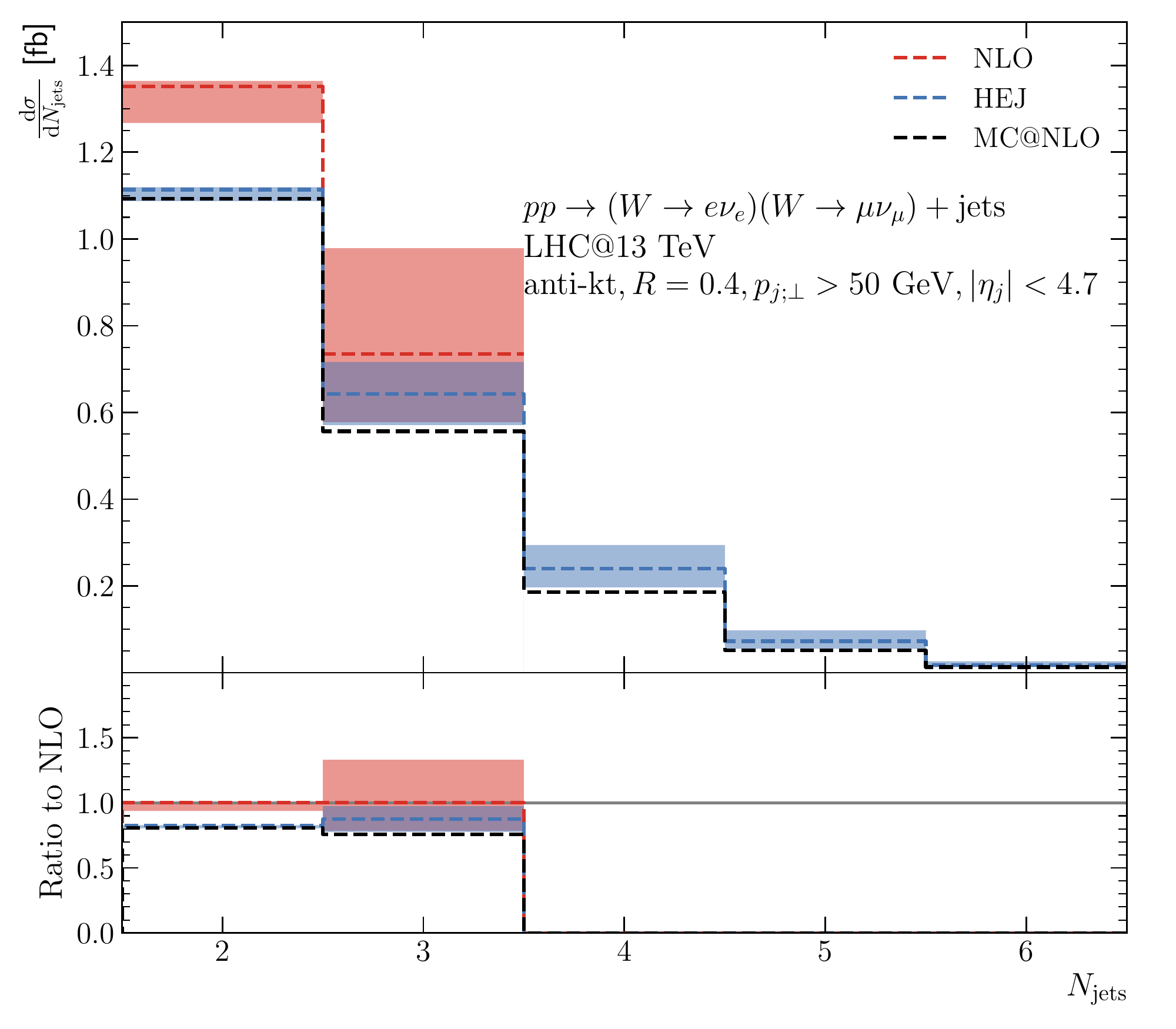}
   \caption{}
   \label{fig:exclusive_jet_rates_novbs}
 \end{subfigure}
 \begin{subfigure}{0.49\textwidth}
   \includegraphics[width=\textwidth]{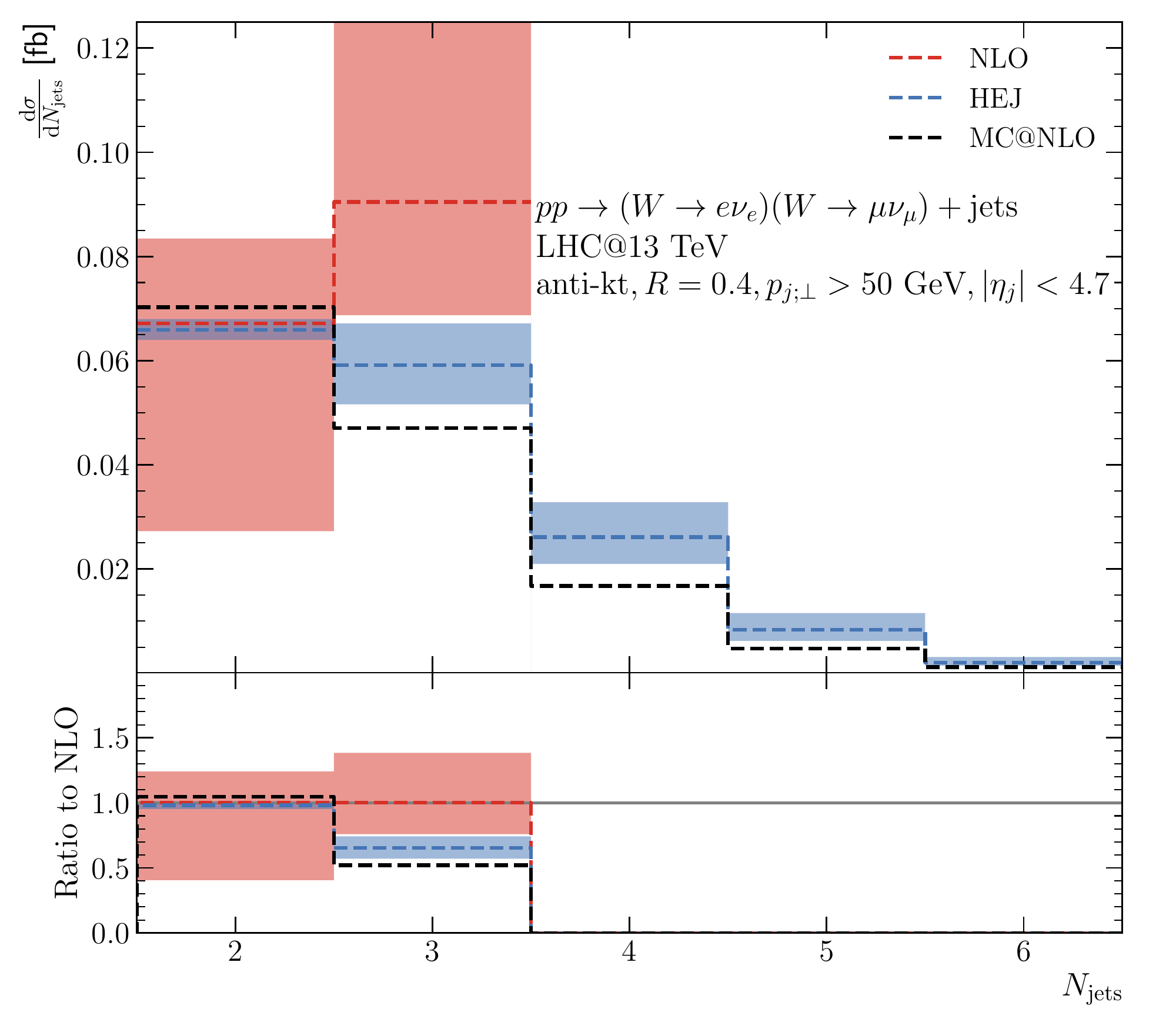}
   \caption{}
   \label{fig:exclusive_jet_rates_vbs}
 \end{subfigure}
\caption{Exclusive jet rates for $pp\to W^\pm W^\pm +\ge2j$, (a) without and (b)
  with additional VBS cuts.}
 \label{fig:exclusive_jet_rates}
\end{figure}

Before discussing the distributions, we give the cross sections obtained at NLO,
with MC@NLO and with \HEJtwo before and after the application of VBS cuts in
Table~\ref{tab:crosssections} for the central scale choice above.  Both before and after VBS cuts the values
are remarkably similar for the central value of the renormalisation and
factorisation scales.  This is a marked difference to other processes where
similar cuts have been applied.  For example in $pp\to H+\ge2j$ despite
relative agreement at the inclusive level, the \HEJtwo predictions were
significantly more suppressed by VBF cuts than those at NLO (by about a factor of
2)~\cite{Andersen:2018kjg}.  This result is sensitive to the scale choice; in the window
of variations we studied, the ratio between the \HEJtwo and NLO cross sections
varies by as much as 22\% in either direction.  It is also clear from the distributions that follow
that this agreement is not flat in phase space (even for the central scale choice), but arises from different
regions where the \HEJtwo result is greater and less than NLO.
\begin{table}[bp]
  \centering
  \begin{tabular}{|c|c|c|c|}
    \hline
    Cross Section (fb)& \emph{without} VBS cuts, $\sigma_{\rm incl}$ & \emph{with} VBS cuts, $\sigma_{\rm VBS} $ &
            $\sigma_{\rm VBS}$/$\sigma_{\rm incl}$   \\ \hline
 \HEJtwo $W^+W^+$ &$1.428 \pm 0.002 $&$0.1219 \pm 0.0004 $&$0.0854 \pm 0.0003 $\\
 NLO $W^+W^+$ &$1.41 \pm 0.05 $&$ 0.12 \pm 0.07 $&$0.08 \pm 0.02 $\\
 MC@NLO $W^+W^+$ &$1.285 \pm 0.003 $&$ 0.1033 \pm 0.0006 $&$0.0804 \pm 0.0005 $\\ \hline
 \HEJtwo $W^-W^-$ &$0.6586 \pm 0.0003 $&$0.0402\pm 0.0001 $&$0.0610 \pm 0.0002 $\\
 NLO $W^-W^-$ &$0.68 \pm 0.02 $&$ 0.04 \pm 0.01 $&$0.06 \pm 0.02 $\\
 MC@NLO $W^-W^-$ &$0.6186 \pm 0.0004 $&$ 0.0371 \pm 0.0002 $&$0.0600 \pm 0.0002 $\\ \hline
  \end{tabular}
  \caption{This table gives the total cross section calculated with the new
    \HEJtwo LO+LL predictions in this paper compared to the result at NLO accuracy, both before and
    after the VBS cuts given in the text.}
  \label{tab:crosssections}
\end{table}
In figure~\ref{fig:exclusive_jet_rates}, we show the exclusive jet rates.  For
only the inclusive selection criteria, we see a steady decrease at each
multiplicity, but in the \HEJtwo predictions the $4$-, $5$- and $6$-jet rates remain at 21\%, 6\% and 2\% of the
exclusive $2$-jet rate respectively. After VBS cuts, the relative importance of
the higher multiplicity rates in the \HEJtwo predictions is enhanced with the $2$-jet and $3$-jet rates being
very similar and the $4$-, $5$- and $6$-jet rates now increasing to 40\%, 13\% and 3\% of the
exclusive $2$-jet rate respectively. In the NLO sample after VBS cuts, the $3$-jet
rate is a third larger than the $2$-jet rate for the central scale choice.  The
scale variation bands here are very large; however, for any one choice the $3$-jet
was always comparable to or greater than the $2$-jet rate.  This is already one
measure of the importance of the higher-order corrections in $\alpha_s$
(i.e.~$\alpha_s^4$ and above).  The MC@NLO predictions shows that the effect of
adding a parton shower to the NLO
predictions is to distribute the $3$-jet component among the higher jet-rate
bins. The large contribution to the cross
section from events with three or more jets suggests that additional
jet vetoes could be used to further suppress the relative QCD
contribution to $pp \to VV+2j$. % Since jet vetoes have not been applied
% in experimental analyses so far, we do not explore this possibility
% any further.

The jet rate plots are affected by the
lepton isolation cut (see table~\ref{tab:cutsCMS}).  Any jet which satisfies
$\Delta R(l,\textrm{jet})<0.4$ for any charged lepton is removed from the event, but the event is still
kept provided there are at least two further jets.  This means that events which
arise from a theoretical calculation with e.g.~4 jets can appear in the plot in
the 2-jet or 3-jet bin.  For comparison, we show the equivalent plots from \HEJtwo without lepton isolation
applied in appendix~\ref{sec:exclusive-jet-rates}.

Figure~\ref{fig:deltayjj} shows
the comparison for the difference in pseudorapidity between the two leading
jets\footnote{We use pseudorapidity to match the convention in the
  Ref.~\cite{Sirunyan:2020gyx}.  In practice, there is little difference in NLO
  and \HEJtwo predictions when one chooses to use rapidity or pseudorapidity as
  the parton multiplicity within a jet is relatively low in each case.}.  Here
we see only modest differences in shape between the two descriptions which are
slightly enhanced once VBS cuts are applied in the right-hand plot.  However,
any differences lie mostly within the scale variation bands.  Here, and in the
remaining distributions, we observe only a small impact of adding a parton shower
to the pure NLO calculation, with no significant changes in shapes of distributions.

Figure~\ref{fig:jet1_pt}  exhibits greater differences in shape in the
distribution of the transverse momentum of the leading jet.  For the
inclusive cuts in (a), the \HEJtwo prediction starts much lower than the NLO
prediction but increases with respect to it until the predictions cross around
200~GeV.  Above this value the prediction from \HEJtwo falls more slowly leading
to a prediction of a harder spectrum in $p_{j_1;\perp}$. A very similar behaviour
is seen in (b) after the application of VBS cuts.  This distribution clearly
emphasises that the close agreement of the total cross section values is a
coincidence of the experimental setup used.  If the transverse momentum
requirement of the jets had been larger, the \HEJtwo cross section would have
also been correspondingly larger than that from NLO.

Similarly, in figure~\ref{fig:m_j1j2} we see that the \HEJtwo and NLO predictions
for the invariant mass distribution of the two leading jets
have a different shape with a ratio which increases steadily from 0.5 at
$m_{j_1j_2}=0$~GeV to 1.4 by $m_{j_1j_2}=2$~TeV where it roughly plateaus.  A similar
effect is seen after VBS cuts are imposed, although of course here the lower
region has been removed.  The point where the predictions cross has moved to a
slightly higher value of $m_{j_1j_2}$ as a result of the cuts on the other variables
which form part of the VBS cuts.

In figure~\ref{fig:m_l1l2}, we show the distributions in the invariant mass of the
two charged leptons from the decays of the $W$ bosons.  This is related to the invariant
mass of the jets if one considers event topologies where the $W$ bosons follow
the direction of the associated quark lines.  For modest transverse momenta, the
invariant mass between particles is driven by their rapidity difference.
The leptons, though, are required to be more central than the jets and we see
more modest differences between the NLO and \HEJtwo predictions.

The final distribution we show in this section is the Zeppenfeld variable $z_e$ of the electron, defined in Eq.~\eqref{eq:zepp}. This measures the relative position of the electron in these
events with
respect to the jet system.  Both before and after VBS cuts, the predictions from
NLO and from \HEJtwo are in very close agreement, and the ratio between the two
remains largely flat throughout the region showing that this variable is largely
insensitive to the logarithmic corrections at higher orders in $\alpha_s$.

In this section, we have compared the new predictions for $pp\to e^\pm \nu_e \mu^\pm
  \nu_\mu+\ge2j$ available in \HEJtwo (which include the leading logarithmic
corrections in $\hat{s}/p_\perp^2$ at all orders in $\alpha_s$) with those obtained at
next-to-leading order in QCD.  We have seen close agreement in the total cross
sections obtained in the two approaches at the central scale choice, but a study of distributions in
$p_{j_1;\perp}$, $m_{j_1j_2}$ and $m_{ll}$ show large differences in shape which
make this agreement appear to be a coincidence of the specific values chosen in the
experimental cuts.

%%% Local Variables:
%%% mode: latex
%%% TeX-master: "main"
%%% End:

\section{Conclusions}
\label{sec:conclusions}

We have presented the calculation of all leading logarithmic contributions which
scale as
\\$\alpha_W^4 \alpha_s^{k+2}\log^k(\hat s/p_t^2)$ to the production of a same-sign
$W$-pair which decays leptonically, i.e.~the QCD contribution to the process $pp\to e^\pm \nu_e \mu^\pm
  \nu_\mu+\ge2j$.
% We have implemented this in the framework
% of \HEJtwo, which has allowed us to additionally match the $n$-jet components to
% leading order accuracy.
In order to separate the electroweak and QCD contributions to this process,
so-called VBS cuts are usually applied to require large invariant mass and
rapidity separation of the tagging jets.  These cuts exactly select regions of
phase space where the logarithms above become important.  The equivalent
corrections have been seen to be significant in the QCD component of
$pp\to H+\ge 2j$ through vector boson fusion, where similar cuts are used. To
assess their impact in $pp\to e^\pm \nu_e \mu^\pm \nu_\mu+\ge2j$, we have compared our new
predictions to those obtained at NLO within the experimental setup of a recent
13 TeV CMS analysis.

We have found that the \HEJtwo cross section is very close to the NLO prediction both for inclusive cuts, and after VBS cuts have been
applied.  However, it is clear from the distributions that this agreement arises
from cancellations across phase space rather than being true throughout.  The
distributions in transverse momentum of the leading jet in
figure~\ref{fig:jet1_pt}, in invariant mass of the leading jets in
figure~\ref{fig:m_j1j2} and in invariant mass of the charged leptons in
figure~\ref{fig:m_l1l2} show clear differences in shape with differences of up to
50\% between \HEJtwo and NLO.  There are other distributions, $\Delta \eta_{j_1j_2}$ and $z_e$
where the two sets of predictions show close agreement, indicating that these
distributions are more stable with respect to higher-order logarithmic corrections.

Previous studies of this process have seen that the $3$-jet component is significant
in typical experimental analyses, enhanced within VBS cuts.  We also find this, and that it extends beyond
$3$-jets.  The exclusive jet components within \HEJtwo are matched to leading-order
accuracy for $2$--$6$~jets.  We showed in figure~\ref{fig:exclusive_jet_rates} that
the VBS cuts do indeed increase the significance of the $3$--$6$-jet components
relative to the $2$-jet component.  The contribution from $3$-jets is similar to
$2$-jet in \HEJtwo and greater than the $2$-jet at NLO.  The $4$--$6$-jet
components steadily decrease but such that the $6$-jet components still
contributed at the order of a few percent in some distributions.

We therefore conclude that logarithmic corrections of the form $\alpha_W^4 \alpha_s^{k+2}\log^k(\hat
s/p_t^2)$ are numerically significant at the 13~TeV LHC, and should be included
in accurate modelling of the QCD background to vector boson scattering.

\section*{Acknowledgements}
\label{sec:acknowledgements}

We are grateful to our collaborators within \HEJ for useful discussions
throughout this project. The predictions presented in section \ref{sec:impact-ll-corr} were produced using resources from PhenoGrid which is part of the GridPP Collaboration\cite{gridpp2006,gridpp2009}. We are pleased to acknowledge funding from the UK
Science and Technology Facilities Council, the Royal Society, the ERC
Starting Grant 715049 ``QCDforfuture'' and the Marie
Sk{\l}odowska-Curie Innovative Training Network MCnetITN3 (grant
agreement no.~722104).

%%% Local Variables:
%%% mode: latex
%%% TeX-master: "main"
%%% End:

\newpage
\appendix
\section{Momentum Configurations for Phase Space Slices}
\label{sec:moment-conf-phase}
We give here the momentum configurations used for the plots of the matrix
elements in figure~\ref{fig:explorers}. % For the 2 jet final state we use:
%
% \begin{equation}
% \begin{split}
% p_i &= p_{i;\perp}(\cos(\phi_i), \sin(\phi_i), \sinh(y_i), \cosh(y_i))\\
% p_{1;\perp} &= p_{\ell;\perp} = 40 \textrm{ GeV} \\
% p_{\nu_\ell;\perp} &=\frac{m_{W}^{2}}{2 p_{\ell;\perp}\left(\cosh \left(y_{\ell}-y_{\nu_\ell}\right)-\cos \left(\phi_{\ell}-\phi_{\nu_\ell}\right)\right)}\\
% p_{2;\perp} &= -(p_{1;\perp} + p_{e;\perp} + p_{\nu_e;\perp} + p_{\mu;\perp} + p_{\nu_\mu;\perp})\\
% \phi_1 &= \pi \quad \phi_e = \pi/4 \quad \phi_\mu = -\pi/2 \quad \phi_{\nu_e} = -\pi/4 \quad \phi_{\nu_\mu} = +\pi/2\\
% y_1 &= y_e = y_{\nu_e} = \Delta \quad y_2 = y_\mu = y_{\nu_\mu} = -\Delta
% \end{split}
% \end{equation}
%
% Whilst for
For the 3 jet final state in figure~\ref{fig:explorers}(a), we use:
\begin{equation}
\begin{split}
p_i &= p_{i;\perp}(\cos(\phi_i), \sin(\phi_i), \sinh(y_i), \cosh(y_i))\\
p_{d;\perp} &= p_{\ell;\perp} = p_{g;\perp} = 40 \textrm{ GeV}\\
p_{\nu_\ell;\perp} &=\frac{m_{W}^{2}}{2 p_{\ell;\perp}\left(\cosh \left(y_{\ell}-y_{\nu_\ell}\right)-\cos \left(\phi_{\ell}-\phi_{\nu_\ell}\right)\right)}\\
p_{s;\perp} &= -(p_{d;\perp} + p_{e;\perp} + p_{\nu_e;\perp} + p_{\mu;\perp} + p_{\nu_\mu;\perp} + p_{g;\perp})\\
\phi_d &= 2\pi/3, \quad \phi_e = \pi/4, \quad \phi_\mu = -\pi/2, \quad \phi_{\nu_e} = -\pi/4, \quad \phi_{\nu_\mu} = +\pi/2, \quad \phi_g = 0.4\\
y_d &= y_e = y_{\nu_e} = \Delta, \quad y_s = y_\mu = y_{\nu_\mu} = -\Delta, \quad y_g = 0
\end{split}
\end{equation}
For the 4 jet final state in figure~\ref{fig:explorers}(b) we use:
\begin{equation}
\begin{split}
p_i &= p_{i;\perp}(\cos(\phi_i), \sin(\phi_i), \sinh(y_i), \cosh(y_i))\\
p_{d;\perp} &= p_{\ell;\perp} =p_{g_1;\perp} = p_{g_2;\perp} = 40 \textrm{ GeV}\\
p_{\nu_\ell;\perp} &=\frac{m_{W}^{2}}{2 p_{\ell;\perp}\left(\cosh \left(y_{\ell}-y_{\nu_\ell}\right)-\cos \left(\phi_{\ell}-\phi_{\nu_\ell}\right)\right)}\\
p_{s;\perp} &= -(p_{d;\perp} + p_{e;\perp} + p_{\nu_e;\perp} + p_{\mu;\perp} + p_{\nu_\mu;\perp} + p_{g_1;\perp} + p_{g_2;\perp})\\
\phi_d &= \pi, \quad \phi_e = \pi/4, \quad \phi_\mu = -\pi/2, \quad \phi_{\nu_e} = -\pi/4, \quad \phi_{\nu_\mu} = +\pi/2, \quad \phi_{g_1} = \pi / 2, \quad \phi_{g_2} = -\pi / 3\\
y_d &= y_e = y_{\nu_e} = \Delta, \quad y_s = y_\mu = y_{\nu_\mu} = -\Delta, \quad y_{g_1} = \Delta  / 3, \quad y_{g_2} = - \Delta  / 3
\end{split}
\end{equation}
The qualitative effects seen in figure~\ref{fig:explorers} are only sensitive to
the rapidity values (and not to the exact choices of transverse momenta or
azimuthal angle).

\section{Simulation Parameters}
In Table~\ref{tab:params} we summarise the parameters used to generate the predictions in section~\ref{sec:impact-ll-corr}.
\begin{table}[H]
  \centering
  \begin{tabularx}{\linewidth}{l|l}
\textbf{Variable} & \textbf{Value}\\
\hline
PDF & NNPDF30\_nlo\_as\_0118 (lhapdf number 260000)\\
$m_Z$ & 91.1876 GeV\\
$\Gamma_Z$ & 2.4952 GeV\\
$m_W$ & 80.385 GeV\\
$\Gamma_W$ & 2.085 GeV \\
$\mu_F$ & $\sqrt{p_{\perp;j_1}p_{\perp;j_2}}$\\
$\mu_R$ &  $\sqrt{p_{\perp;j_1}p_{\perp;j_2}}$\\
$\alpha_s(m_Z)$ & 0.118 (set to match PDF)\\
$1/\alpha_w$ & 132.232\\
$G_F$ & $1.16639\times 10^{-5}\ \textrm{GeV}^{-2}$ \\
Sherpa Version & 2.2.2\\
Openloops Version & 1.3.1\\
Parton Shower & CS Shower (packaged with Sherpa)
\end{tabularx}
\caption{The input parameters used for the plots in section \ref{sec:impact-ll-corr}.}
\label{tab:params}
\end{table}
Sherpa uses the complex mass scheme for vector boson masses and $\alpha_w$ is calculated in the $G_\mu$ scheme from these.

%%% Local Variables:
%%% mode: latex
%%% TeX-master: "main"
%%% End:

\section{Exclusive Jet Rates Without Lepton Isolation Requirement}
\begin{figure}[t]
\centering
 \begin{subfigure}{0.49\textwidth}
   \includegraphics[width=\textwidth]{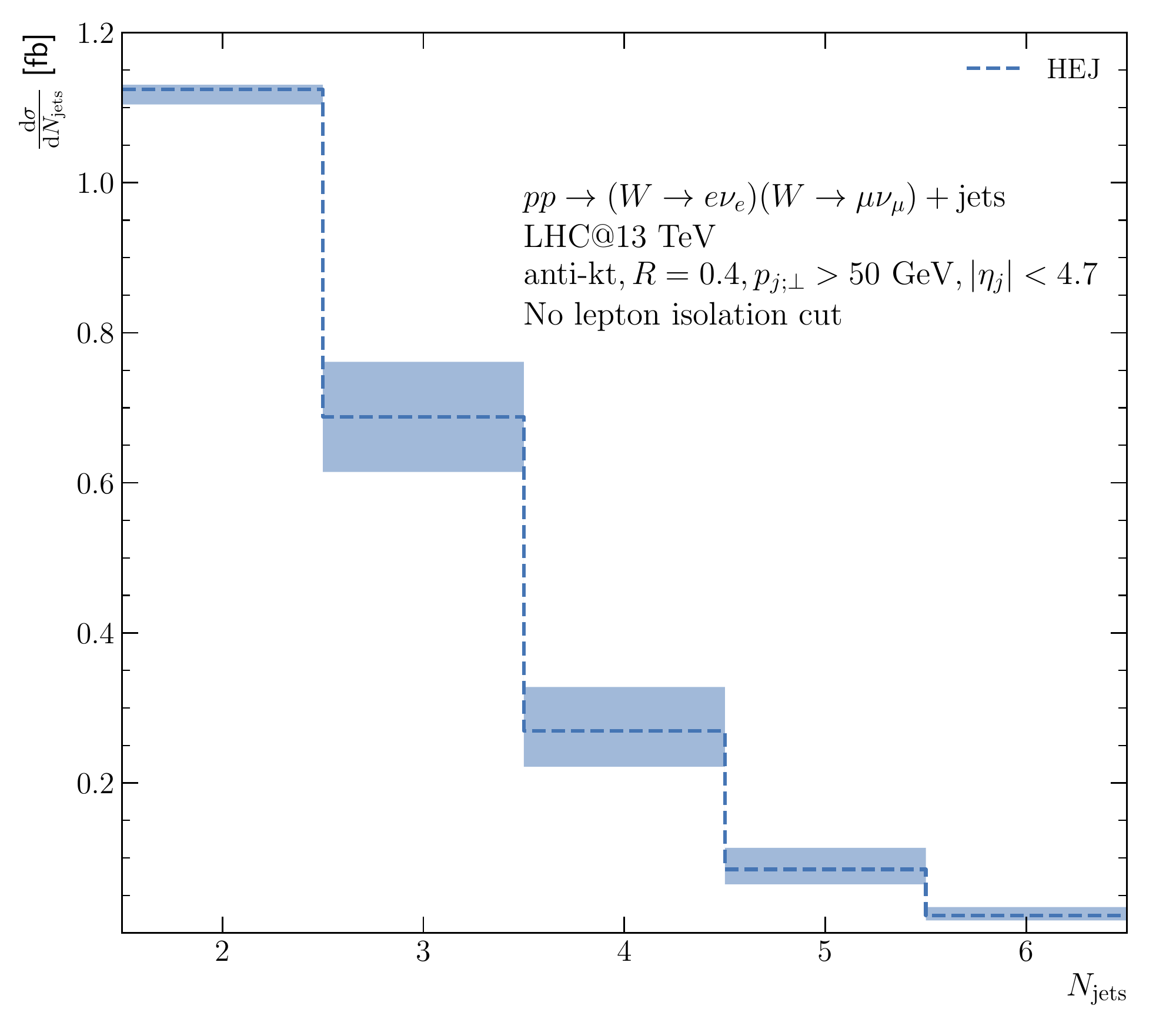}
   \caption{}
   \label{fig:no_isolation_jet_rates_novbs}
 \end{subfigure}
 \begin{subfigure}{0.49\textwidth}
   \includegraphics[width=\textwidth]{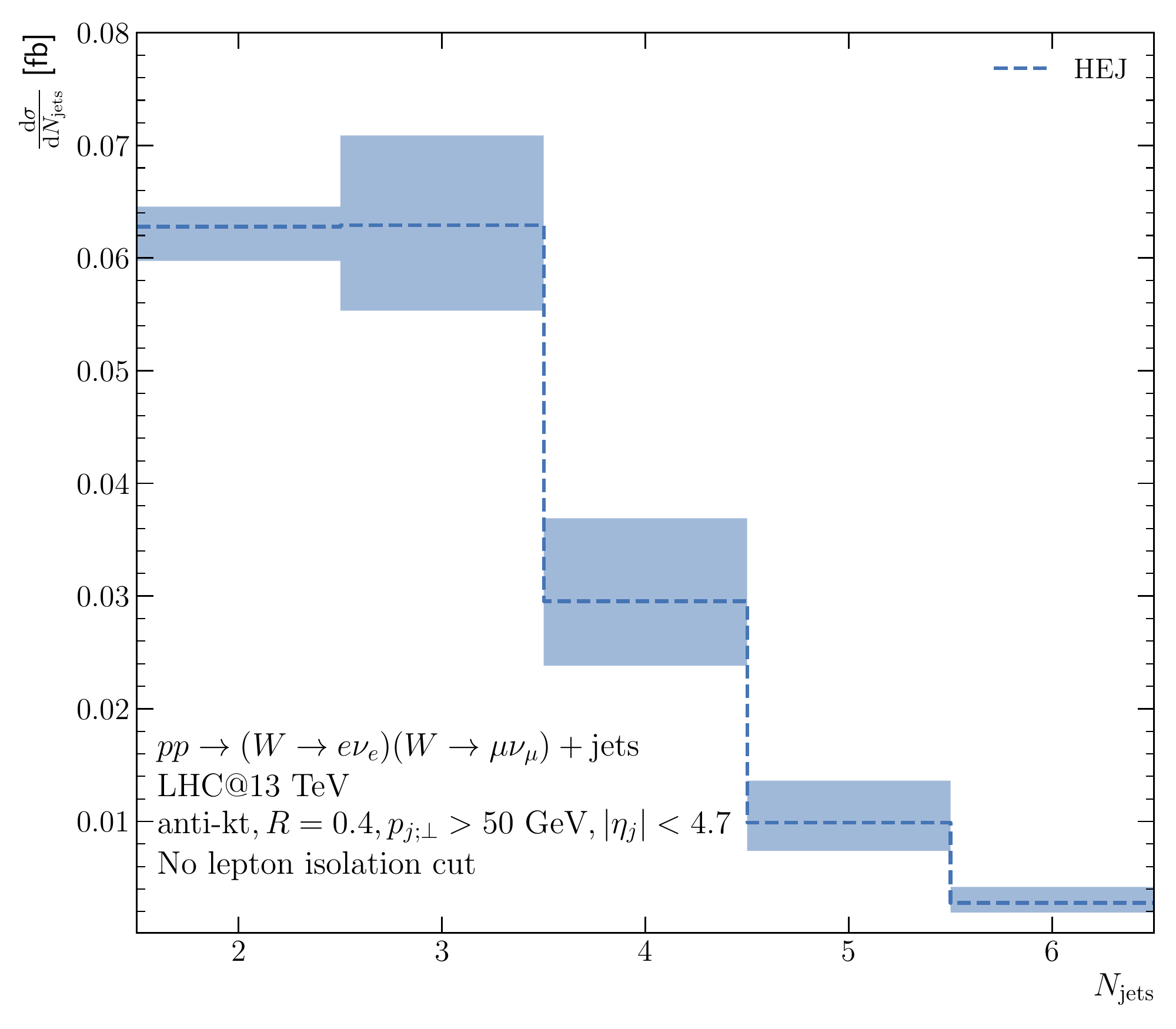}
   \caption{}
   \label{fig:no_isolation_jet_rates_vbs}
 \end{subfigure}
 \caption{Exclusive jet rates for $pp\to W^\pm W^\pm +\ge2j$ as in
   figure~\ref{fig:exclusive_jet_rates} except now the lepton isolation
   requirement has been
   removed, (a) without and (b) with additional VBS cuts.}
 \label{fig:no_isolation_jet_rates}
\end{figure}

In figure~\ref{fig:exclusive_jet_rates} we showed the exclusive jet rates obtained
at NLO and with \HEJtwo for the experimental setup described in
Table~\ref{tab:cutsCMS}.  This includes a lepton isolation requirement where any
jet which satisfies $\Delta R(l,{\rm jet})<0.4$ is removed from the event.  This
means that many events appear in bins with lower numbers of jets than that
present in the calculation of the weight of that event.  To more closely reflect
the impact of the higher orders in the calculation, in
figure~\ref{fig:no_isolation_jet_rates} we show the jet rates from \HEJtwo with
the lepton isolation requirement removed (all other cuts remain the same).
\label{sec:exclusive-jet-rates}
The differences with only inclusive cuts are modest, with a slight decrease in
the first bin and slight increases in the higher bins.  After VBS cuts, the
effect is more pronounced.  The $3$-jet rate is now slightly above the $2$-jet rate
and there is then a bigger step down to the $4$, $5$ and $6$ jet rates which
have each risen slightly from the values after lepton isolation cuts.  They are
now 47\%, 16\% and 4\% respectively of the exclusive $2$-jet rate compared to
40\%, 13\% and 3\% after the lepton isolation cut is applied.

%%% Local Variables:
%%% mode: latex
%%% TeX-master: "main"
%%% End:

\bibliographystyle{JHEP}
\bibliography{papers}

\end{document}